\documentclass[%
 aip,
 jmp,%
 amsmath,amssymb,
 reprint,%
]{revtex4-1}

\usepackage{graphicx}
\usepackage{dcolumn}
\usepackage{bm}

\begin{document}

\title[]{The Many-Body Expansion Combined with Neural Networks.}

\newcommand*{\mycommand}[1]{\texttt{\emph{#1}}}
\author{Kun Yao}
\author{John E. Herr}
\affiliation{251 Nieuwland Science Hall, Notre Dame, IN 46556}
\author{John Parkhill}
\affiliation{251 Nieuwland Science Hall, Notre Dame, IN 46556}
\email{john.parkhill@nd.edu}

\date{\today}
             
\begin{abstract}
Fragmentation methods such as the many-body expansion (MBE) are a common strategy to model large systems by partitioning energies into a hierarchy of decreasingly significant contributions. The number of fragments required for chemical accuracy is still prohibitively expensive for ab-initio MBE to compete with force field approximations for applications beyond single-point energies. Alongside the MBE, empirical models of ab-initio potential energy surfaces have improved, especially non-linear models based on neural networks (NN) which can reproduce ab-initio potential energy surfaces rapidly and accurately. Although they are fast, NNs suffer from their own curse of dimensionality; they must be trained on a representative sample of chemical space. In this paper we examine the synergy of the MBE and NN's, and explore their complementarity. The MBE offers a systematic way to treat systems of arbitrary size and intelligently sample chemical space. NN's reduce, by a factor in excess of $10^6$ the computational overhead of the MBE and reproduce the accuracy of ab-initio calculations without specialized force fields. We show they are remarkably general, providing comparable accuracy with drastically different chemical embeddings. To assess this we test a new chemical embedding which can be inverted to predict molecules with desired properties.

\end{abstract}

\keywords{many-body expansion, neural networks, methanol}
\maketitle

\section{Introduction}
\indent  The many body expansion (MBE) lies at the heart of a multitude of computational methods being developed in the realm of ab-initio theory and force fields. In insulators the high-order terms of the MBE decay rapidly with distance, which makes this type of approximation useful for low-scaling, high-accuracy models of liquids, solids and biological molecules\cite{raghavachari2015accurate,mayhall2012many, richard2012generalized,liu2016pair, wen2012practical,beran2009approximating,saha2013dimers,gordon2012fragmentation,dahlke2007electrostatically,medders2013critical,von2010two}.  However an ab-initio MBE is orders of magnitude more costly than a classical force field. The main limitation comes from the combinatorial growth of effort at each order. \\
 \indent In chemistry neural networks are growing in popularity to predict molecular properties\cite{jose2015vibrational,behler2007generalized, behler2011neural,jiang2013permutation,handley2010potential,cuny2016ab,zhang2014effects,koch2014communication, chen2013communication}. However NN's have their own limitations: their input must have a constant shape, moreover they must be trained on a representative number of samples, and chemical space grows exponentially with molecular size. This curse of dimensionality in the training set is the main barrier to the creation of a universal NN force field with very high accuracy. The purpose of this paper is to show that the MBE provides a very natural and accurate way to alleviate this curse of dimensionality while retaining the generality, accuracy and efficiency of a NN.\\
 \indent    Force fields based on the many-body expansion are growing in popularity. \cite{richard2012generalized,dahlke2007electrostatically,beran2009approximating,pinski2013reactive}. Under the MBE scheme, the total energy of an system can be expanded as the sum of the many-body terms. High order terms are more costly calculations and the error of the MBE is often balanced with the error of the underlying model chemistry at third order\cite{xantheas1994abinitio,xantheas2000cooperativity,kulkarni2004manybody} so long as care is taken to correct for basis set superposition error (BSSE).\cite{ouyang2014trouble,boys1970calculation} An electrostatically embedded MBE (EE-MBE) has also been proposed as a means to improve the accuracy.\cite{dahlke2007electrostatically, dahlke2007electrostatically2}. Others have suggested a many-body expansion scheme of overlapping-fragments as a way to improve upon the accuracy of the energies.\cite{richard2012generalized,richard2013many-body,lao2016understanding} \\
\indent Statistical models from machine learning are becoming popular chemical models. Examples include fitted potential energy surfaces\cite{behler2011neural, handley2010potential} with atom-centered symmetry functions,\cite{behler2007generalized,behler2011atomcentered,khaliullin2011nucleation} and with permutation invariant polynomials.\cite{jiang2013permutation,xu2014global,zhang2014effects,medders2013critical, shao2016communication, li2015permutationally, li2015permutationally2,medders2015representation} Permutationally invariant polynomials have been used to express the many-body energies of water clusters\cite{medders2013critical, medders2015representation} and water-methane clusters\cite{conte2015permutationally} with great success. Also, machine learning has been used to predict properties, such as atomization energies, HOMO and LUMO eigenvalues, ionization potentials, force constants, dielectric constants \cite{Rupp:2015aa,Hansen:2015aa,montavon2013machine,pilania2013accelerating, ghasemi2015interatomic, schutt2014represent, olivares2011accelerated,ma2015machine, ediz2009using}, quantum transport coefficients\cite{lopez2014modeling} and nuclear magnetic resonance parameters\cite{cuny2016ab}. It has also been used to construct kinetic energy functionals\cite{Snyder:2013aa, snyder2012finding, yao2016kinetic} and to design new materials\cite{olivares2011accelerated, hachmann2011harvard, hachmann2014lead, hautier2010finding, ediz2008molecular}. \\ 
\indent To our knowledge, there are few works that combine neural networks with the MBE and those have focused on elemental solids. The closest work predicted the many-body energy of $Si_{n} (n=1,2,...7)$ \cite{malshe2009development} clusters. Bart{\'o}k used machine learning techniques based on Bayesian inference to correct DFT one-body and two-body energies for water\cite{bartok2013machine}.  In this paper we learn the many-body energies of condensed phase liquid methanol within m$E_h$ accuracy. We show that one can use the MBE for methanol clusters of a thousand molecules without significant computational expense on typical GPU workstations. We also present a novel chemical embedding, which has the advantage that it is invertible to ball-and-stick geometries, asses it as a descriptor to learn the MBE, and propose it as a useful tool for inverse-design. \\
 
\section{Methods}

\indent Studies have shown that the MBE converges rapidly for van der Waals and water clusters.\cite{cui2006theoretical,gora2011interaction,ouyang2014trouble,hermann2007convergence,medders2013many} Convergence is relatively slow for metallic or covalent interactions\cite{hermann2007convergence,paulus2004convergence}, although schemes have been proposed to improve the accuracy of the MBE on covalent systems\cite{mayhall2012many,richard2012generalized}. We chose methanol for its strong hydrogen bonding, but nothing about this work is specialized or limited to systems of this size. RI-MP2 with the cc-pVTZ basis is used to calculate all of the training and testing data for the many-body energies. The integral precision and SCF convergence criteria were as tight as possible\cite{richard2014aiming,richard2014understanding} and BSSE using the K-mer centered basis set approach ($k-CBS$)\cite{gora2011interaction} was applied. Training and test geometries are drawn from an AMBER molecular dynamics trajectory\cite{fennell2006ewald,lamichhane2014real1,lamichhane2014real2,Amber} of 108 methanol molecules at 330 K and ab-initio trajectory of three methanols at 500 K. The total data set include 844,800 samples one-body energies, 74,240 samples for two-body energies and 36,864 samples for three-body energies. $20 \%$ of the total data set is used for testing. All of the ab-initio calculations are done with the $Q-Chem$ package \cite{shao2015advances}. Previous studies have shown that cumulative two-body energies and cumulative three-body energies converged at a cutoff of 10 \AA \ for $(H_2O)_{21}$.\cite{cui2006theoretical, dahlke2007electrostatically2}. We also found that both the two-body and three-body energies negligibly different from limiting formulas at a cutoff of 10 \AA \ as shown in Figure SI-1 so our dimers and trimers were generated within this cutoff of 10 \AA. $Cuda-Convnet$\cite{krizhevsky2012imagenet} was used to train and evaluate the neural network. \\

\begin{figure}[t]
\includegraphics[width=0.5\textwidth]{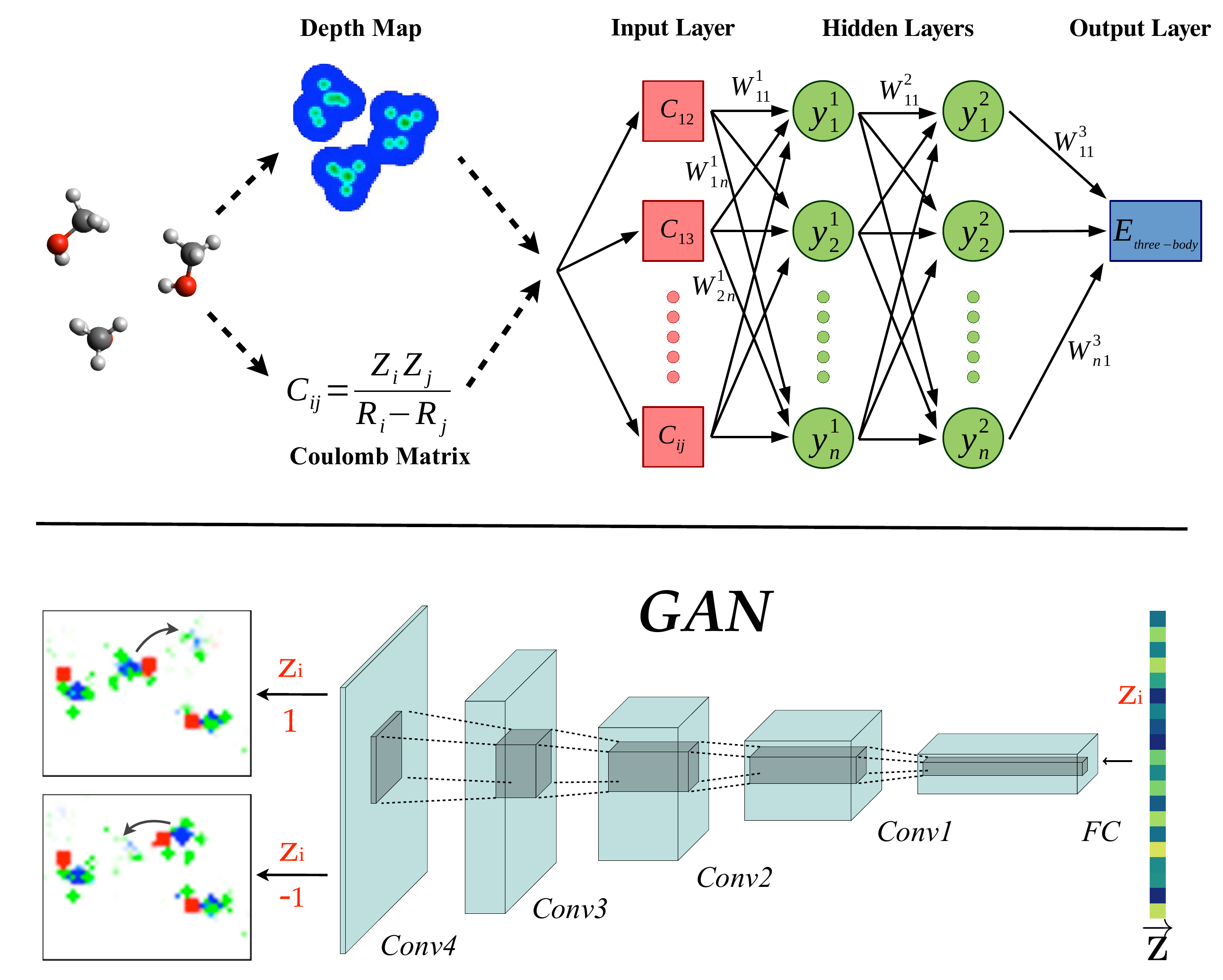}
\caption{Top panel: The fragment energy of each N-Body is calculated by embedding the geometry into either the Coulomb Matrix or Depth map, and evaluating the output of a neural net with several hidden layers and one output. Bottom panel: Generative adversarial network scheme. A z-vector is transformed and passed through convolutional hidden layers to generate a hallucinated depth map.}
\label{NNscheme.pdf}
\end{figure}

\indent Choosing the chemical embedding for the system as the input to the NN has a great effect on performance. Many different chemical descriptors have been proposed, including the Coulomb matrix\cite{hansen2013assessment, montavon2012learning, faber2015crystal}, symmetry functions\cite{behler2007generalized,behler2011neural}, bispectrum\cite{bartok2010gaussian, bartok2013representing}, permutation invariant polynomials\cite{zhang2014effects, jiang2013permutation}, metric fingerprints\cite{sadeghi2013metrics, zhu2016fingerprint,SchaeferComputationally} and  the radial distribution Fourier series, \cite{von2015fourier} which is based on the electronic density and is similar to a descriptor our group has used in the past for learning kinetic functionals\cite{yao2016kinetic}. Systematic comparison of different descriptors is beyond the scope of this paper and we choose the Coulomb matrix (CM) as our input for neural networks for its simplicity and we show that it is capable of the task. The CM, however, is not permutationally invariant, therefore, in this study we augmented our training data with all the permutations of hydrogen atoms on carbon and all the permutations of methanol molecules in the dimer and trimer to learn the permutation invariance. Similar data augmentation techniques have been widely used in image recognition to achieve translation and rotation invariance\cite{krizhevsky2012imagenet}. As shown in the Figure SI-2 and Figure SI-3, the permutation invariance is learned with satisfying accuracy. The permutation invariance can also be avoided by averaging the result of all the possible permutations.\\ 
\indent We experimented with a novel chemical embedding, which we call the depth map (D-map). The purpose of this descriptor is not to improve over the accuracy of the CM, but rather to have an input which provides reasonable energies and inverts directly to molecular geometries. If networks could accurately learn from an invertible input, they could also become useful tools for molecular inverse-design.  Similar types of NN inputs have been used in the area of 3D detection and object recognition.\cite{song2014sliding,shrivastava2013building,lai2012detection,kim2013accurate} An example D-map can be seen in Figure~\ref{NNscheme.pdf}. It is simply a depth of field image of a ball-and-stick structure. A simple routine to calculate one is given in the supplement. Given the usefulness of this input in molecular design we were curious how well it could be used to predict energies, and will compare it to the CM in the results. Generative Adversarial Networks (GANs) have since been studied extensively for their ability to hallucinate authentic looking images.\cite{goodfellow2014generative,denton2015deep,radford2015unsupervised,im2016generating,yoo2016pixel-level,goodfellow2016improved} We trained a GAN on the D-map to produce hallucinated images of methanols, and discuss the utility this provides.

\section{Results}
\begin{figure}[t]
\includegraphics[width=0.5\textwidth]{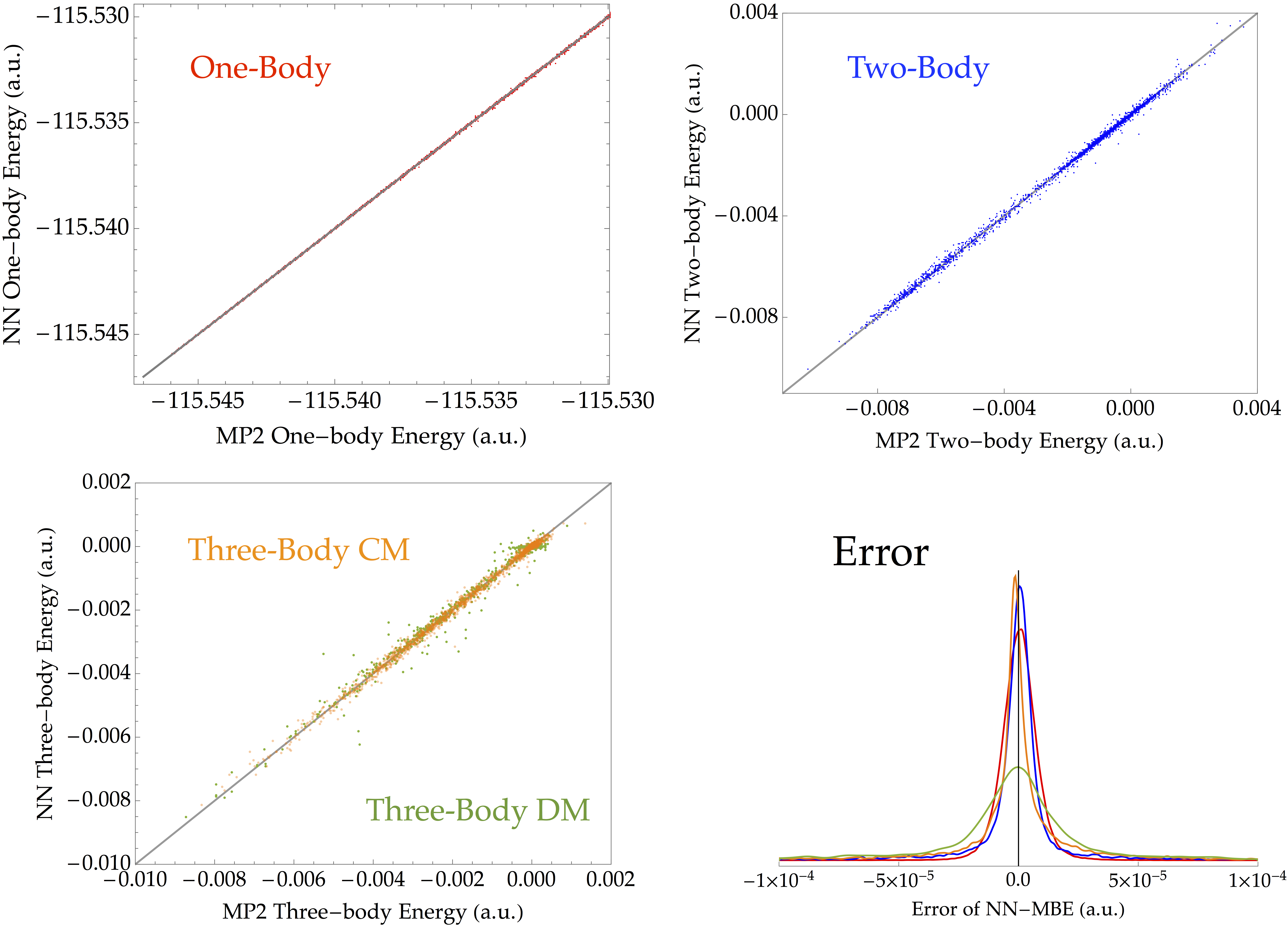}
\caption{ The top left, top right and bottom left panels show plots of the one-body, two-body and three-body energies calculated from MP2 (x-axis) and our neural network (y-axis), respectively. For three-body energy, the result of using the Coulomb matrix (CM) as the input is shown in orange and the result of using the depth map (DM) as the input is shown in green. The bottom right panel shows the histogram of the errors of the many-body energy terms predicted by the neural network. The one-body and two-body errors are shown in red and blue respectively.}
\label{onetwothree_depthmap.png}
\end{figure}

\indent Figure~\ref{onetwothree_depthmap.png} shows the comparison of the one-body, two-body and three-body energies calculated using MP2 and our neural network for all the test samples. The neural networks give close agreement with MP2 such that errors in the underlying model chemistry would be the limiting factor of MBE-NN. The Mean Absolute Error (MAE) and Mean Signed Error (MSE) of the energy of each many-body terms over a test set independent of the training data are shown in Table \ref{tbl:maemse}. The MSE is balanced in the sense that each order of the expansion adds comparable microhartree errors to a total energy, as we will discuss in results which follow. \\
\indent     The higher MAE of the higher-body terms is a predictable consequence of our design principle that the ratio of total error to wall-hours should be kept roughly constant at each order of expansion. The three-body energy is a surface of much higher dimension than the two body energy, and this causes difficulty in three ways: a need for more training data, a larger number of invariances, and network capacity. However, we use much less three body data, because it is more expensive and less significant. The histogram of errors of all the many-body terms are also shown in Figure~\ref{onetwothree_depthmap.png}, and appears uncorrelated which is supported by our later observation that the error per-methanol is essentially insensitive to system size.\\

\begin{table}
  \caption{ The MAE and MSE (microhartree) of one-body energy, two-body energy, three-body energy  with Coulomb matrix input and three-body energy with depth map input predicted by neural network. We calculate a rate of error as MAE(microhartree)/wall-hours of RI-MP2 in Q-Chem required to generate the training data.}
  \label{tbl:maemse}
  \begin{tabular}{lllll}
    \hline
    Error   &  1-body  &  2-body  & 3-body (CM) & 3-body (DM)\\
    \hline
    MSE & 0.24  & 0.90 & -1.16 &  -3.96 \\
    MAE   & 5.99  & 15.6 &  20.0  & 39.0 \\
    Rate & 0.005  & 0.008 &  0.005  & 0.009\\
    \hline
  \end{tabular}
\end{table}

\indent The three-body network trained on the D-map also provides reasonable energies. Comparing the three-body energy plots for the CM and the D-map, in most cases the D-map appears to do nearly as well as the CM; however, for a handful of cases the D-map makes significantly poorer predictions. We note this tends to happen when all or part of an oxygen atom becomes eclipsed by the methyl group. Furthermore, we also note the D-map tends to make poorer predictions for energies which are near zero. The distribution of errors remains normally distributed. The D-map does provide a few advantages over the CM: It provides a low dimensional encoding of the space of methanol geometries. It also has constant shape regardless of chemical input, and suffers from fewer problems with invariances. \\ 
\indent As expected the CM makes more accurate predictions than the D-map; however, our aim with the D-map was to provide a chemical embedding which can easily be mapped back to the original geometry of the system. For example this can be used to predict molecules, which maximize a desired property directly without searching chemical space. To this end we then trained a GAN, based on Radford et al.,\cite{radford2015unsupervised} using the D-map by separating element types into separate color channels, to produce hallucinated images of methanol trimers. An example hallucinated D-map can be seen in the bottom panel of Figure~\ref{NNscheme.pdf}. The network maps a random z-vector back to a D-map. By varying one element of the z-vector, we were able to control image generation to tune properties. The examples in Figures~\ref{NNscheme.pdf} are from the same z-vector with one varied element $z_i$, which rotates one of the methanol end over end. It is easy to imagine extending this to other properties and using a GAN for inverse-design. For the remainder of this paper we employ the Coulomb embedding for the MBE.

\begin{table}
  \caption{Relative energies (m$E_h$) of three minimal energy geometries of methanol trimer. }
  \label{tbl:notes}
  \begin{tabular}{lllll}
    \hline
    Geometry   &  MP2\textsuperscript{\emph{a}}  &  MBE-NN & HF\textsuperscript{\emph{a}} & B3LYP\textsuperscript{\emph{a}}\\
    \hline
    chair   &  0 & 0 & 0  & 0 \\
    bowl & 1.50  & 1.40 & 1.75 &  2.02\\
    chain & 4.54 & 4.10  & 2.56 &  5.29 \\
    \hline
  \end{tabular}

  \textsuperscript{\emph{a}} MP2, HF and B3LYP energies are extrapolated to a complete basis.
\end{table}

\begin{figure}[t]
\includegraphics[width=0.4\textwidth]{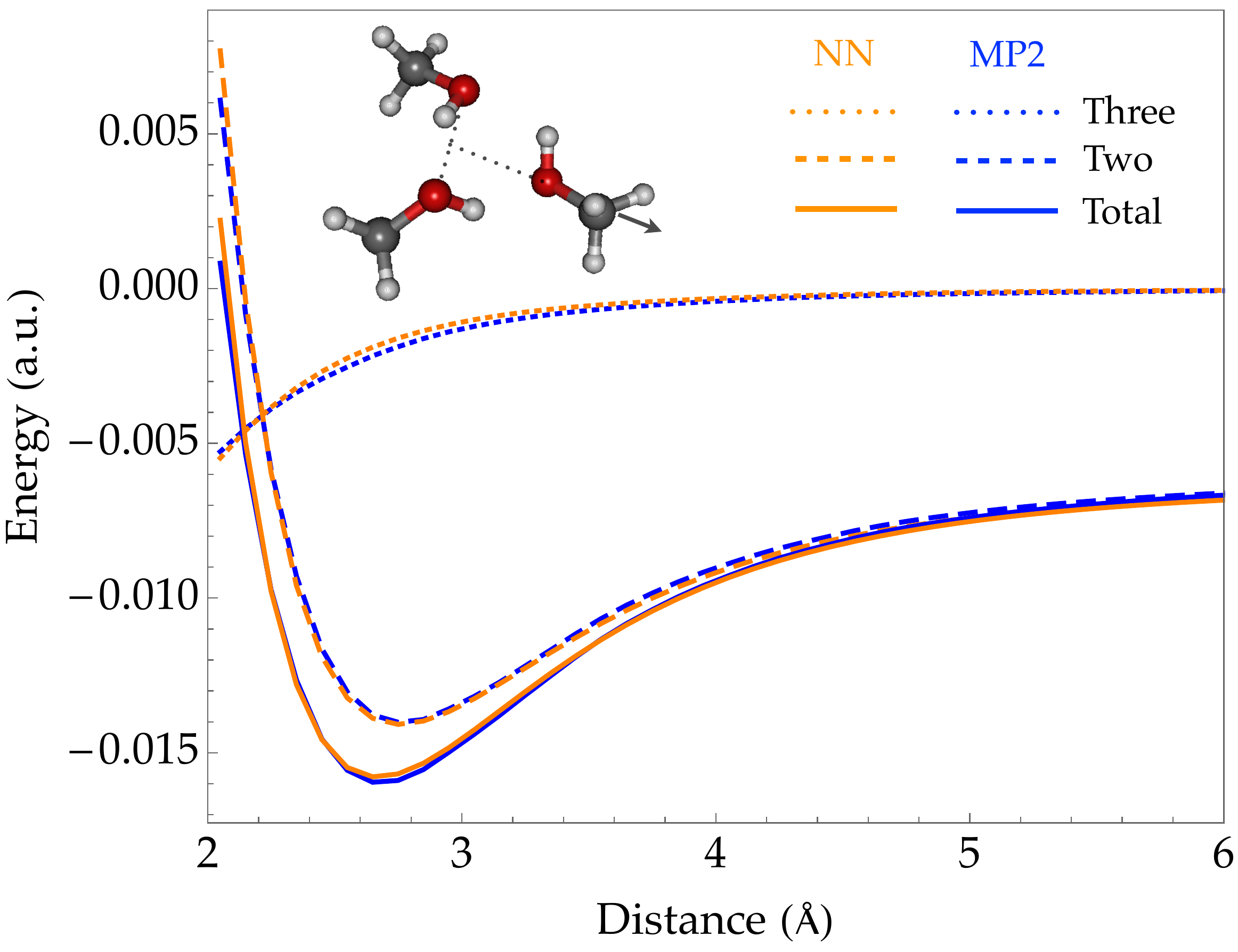}
\caption{ Dashed line, dotted line and solid line show the change of three-body energy, two-body energy and total energy when one methanol is pulled away from the other two. Energies calculated by MP2-MBE and NN-MBE are shown in blue curve and orange curve, respectively.  }
\label{fig:strechbond.pdf}
\end{figure}

\indent  Table~\ref{tbl:notes} gives the relative energies of the three minimal energy geometries, chair, bowl and chain\cite{kazachenko2013methanol}  of a methanol trimer. All methods shown in Table~\ref{tbl:notes} get the ordering of these three geometries correct. Compared with MP2, our many-body expansion neural network (NN-MBE) has an error within $10\%$, which is small compared to Hartree-Fock and B3LYP, which are both significantly more costly. Another important feature is the smoothness of predicted surfaces.  Figure~\ref{fig:strechbond.pdf} plots the change of the three-body energy, two-body energy and total energy when one methanol in a methanol trimer is pulled away from the other two. The agreement is best near the bonding minimum and long distance. Relatively fewer training cases in this region are sampled by the MD trajectories.

\begin{figure}[t]
\includegraphics[width=0.4\textwidth]{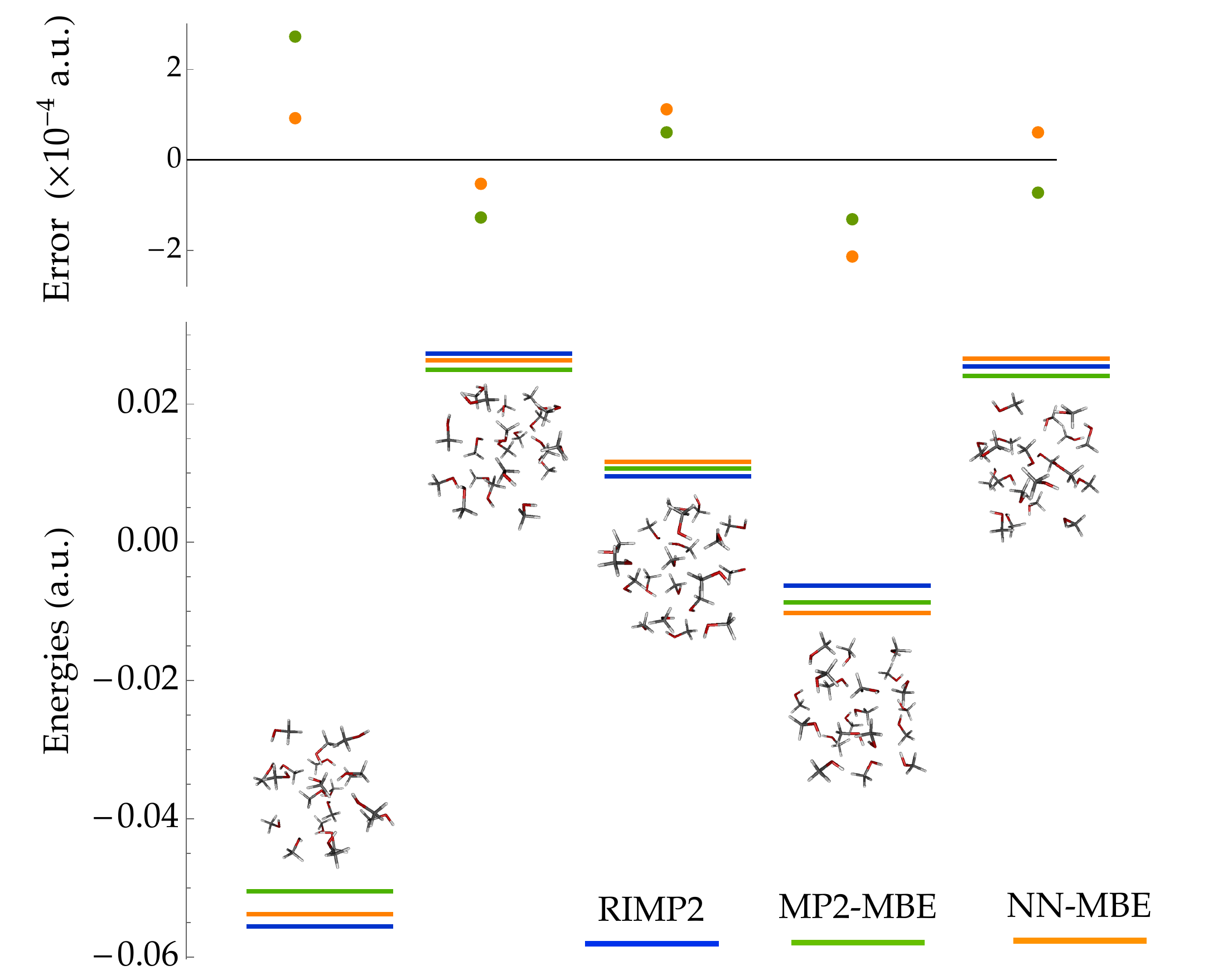}
\caption{ Bottom panel: relative energies of five $\text{MeOH}_{20}$ clusters. Energies calculated by MP2, MP2-MBE and NN-MBE are shown in blue lines, green lines and orange lines, respectively. Top panel: green dots show the difference between MP2-MBE energies and MP2 energies and orange dots show the difference between NN-MBE energies and MP2 energies.}
\label{fig:20cluster.pdf}
\end{figure}

\indent The relative energies of five random $(\text{MeOH})_{20}$ clusters are shown in Figure~\ref{fig:20cluster.pdf} to assess the errors due to NN and MBE. Compared with the real MP2 energies, the MAE of the the MBE using MP2 (MP2-MBE) is 0.12 m$E_h$ per molecule, and with NN-MBE 0.10 m$E_h$. Remarkably there is no degradation of accuracy involved in using NN-MBE, despite massive speedup. Instead the method is limited by the quality of the model chemistry it is built on and the accuracy of the MBE itself.  

\begin{figure}[t]
\includegraphics[width=0.28\textwidth]{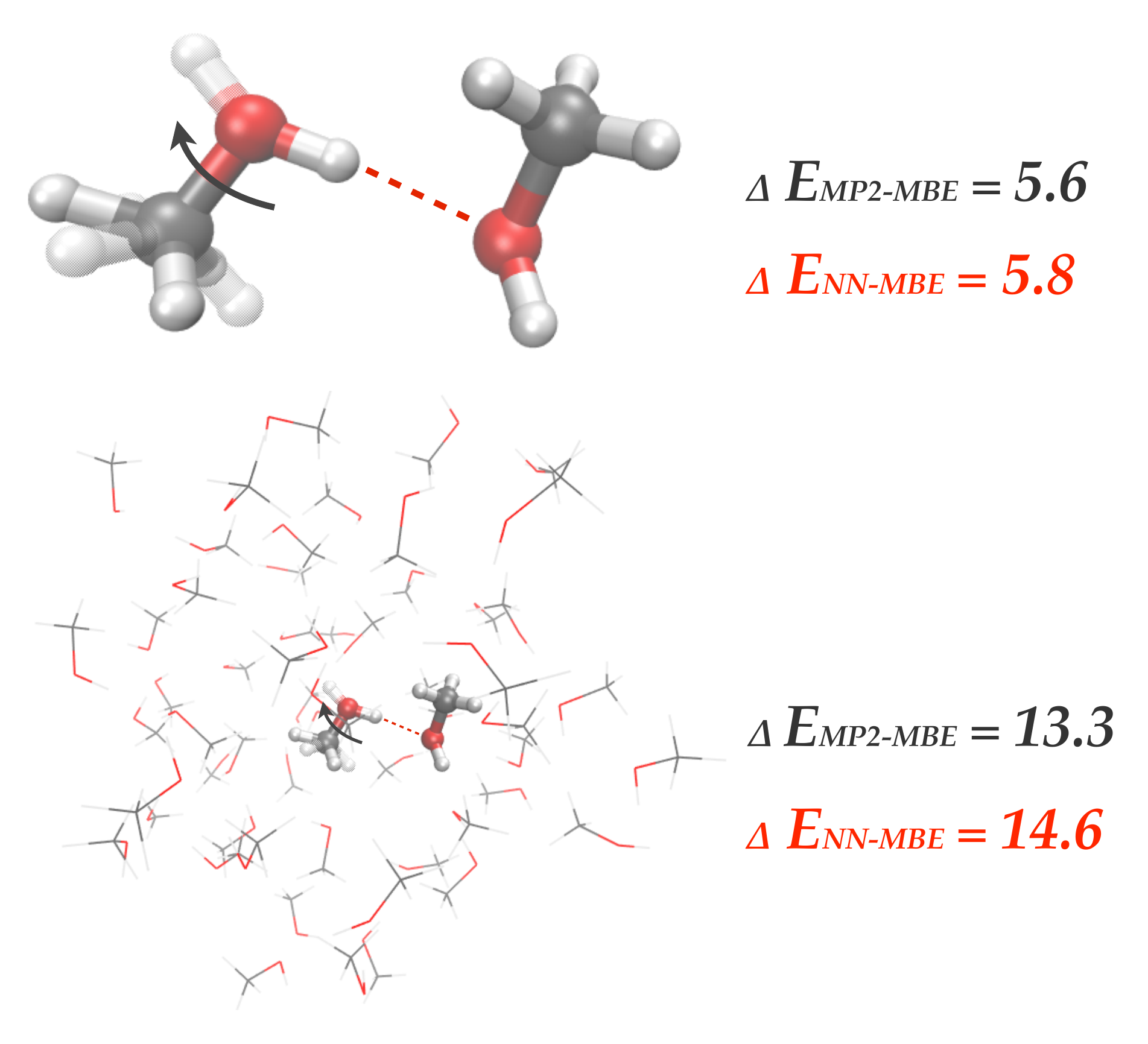}
\caption{ The top panel (without solvation shell) and bottom panel (with solvation shell) show the energy changes of breaking a hydrogen bond between two methanol by rotating one methanol by 180 degrees around the C-O bond. The units of the energy are kcal/mol. The solvation shell influences the energy change significantly and the neural network predicts the energy change with an accuracy of 1 kcal/mol }
\label{fig:flipH_new.pdf}
\end{figure}

\indent Proper treatment of solvent effects is crucial for describing most chemical processes. The top panel of Figure~\ref{fig:flipH_new.pdf} shows the energy change of breaking the hydrogen bond between two methanols when the solvation shell is not included. MP2-MBE predicts the energy change to be 5.6 kcal/mol and NN-MBE gives 5.8 kcal/mol. When the solvation shell (with a radius of  10 \AA) is included, as shown in the bottom panel, the energy change dramatically increased to 13.3 kcal/mol, which shows the large influence of solvent effects. The NN-MBE predicts the energy change with solvation shell to be 14.6 kcal/mol, 1.3 kcal/mol larger than the MP2-MBE result. Considering the speed up of the NN-MBE, discussed below, and its accuracy, the scheme shows promise for condensed phase phenomena.  

\begin{figure}[t]
\includegraphics[width=0.35\textwidth]{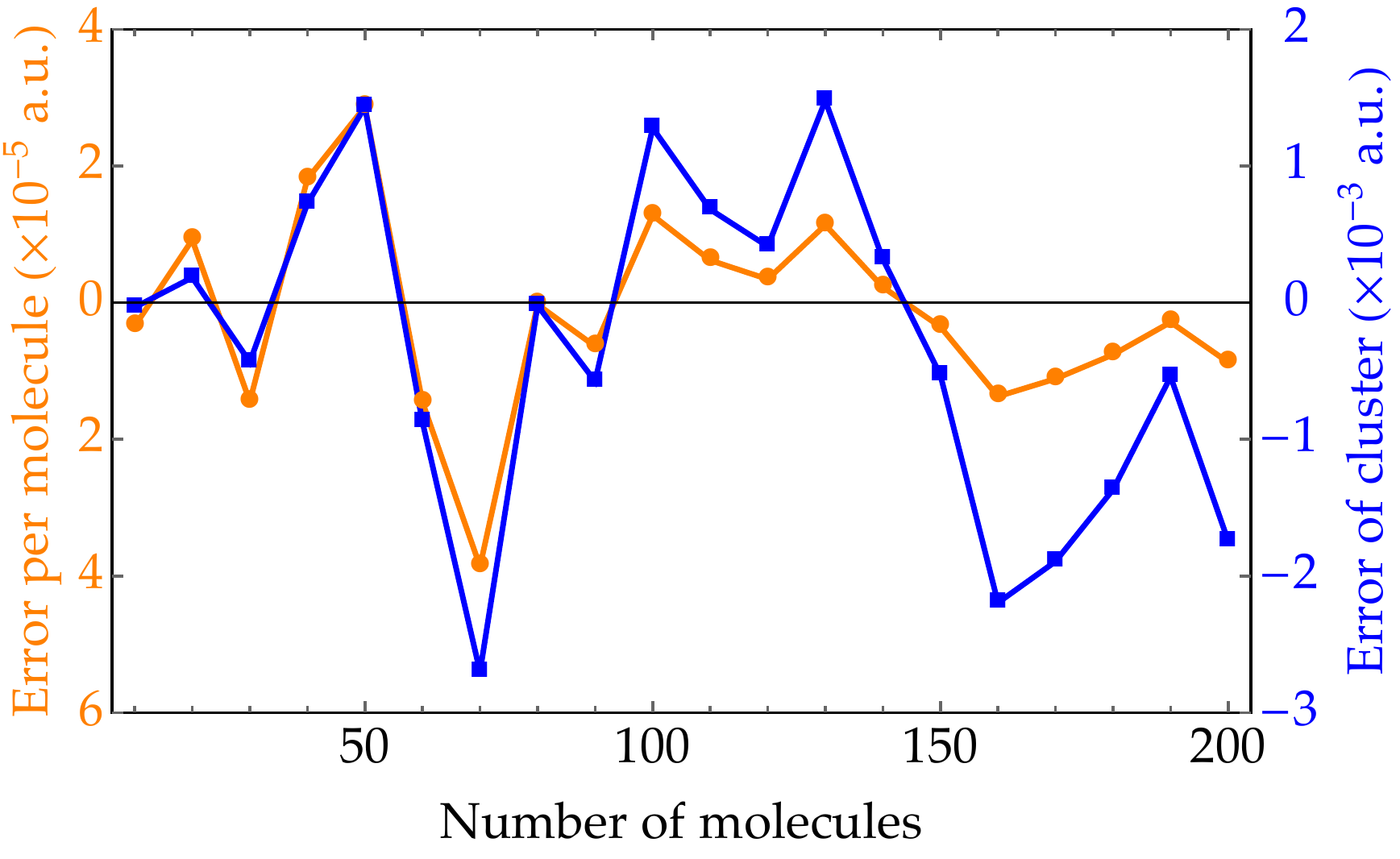}
\caption{Error per MeOH (left orange axis) and total error per cluster (right blue axis) of methanol clusters with different size. Error is define as the difference between NN-MBE energy and MP2-MBE energy. The total error of per cluster which includes up to 200 molecules is in the range of 3 m$E_h$ and the error per molecule decreases with system size.}
\label{200mol.pdf}
\end{figure}

\indent We also investigated the error of NN-MBE as a function of system size. Figure~\ref{200mol.pdf} shows the error per molecule and the error per cluster of the NN-MBE (with respect to MP2-MBE) with an increasing number of molecules in the cluster. The error per cluster stays within the range of 3 m$E_h$ and the error per molecule reaches a maximum at 70 units and shows signs of sub-extensive behavior. Figure SI-5 provides the total wall time comparison of the NN-MBE and MP2-MBE, showing that the NN-MBE offers a speed up of more than \textbf{two million} relative to MP2-MBE without any type of optimization.

\section{Discussion and Conclusions}

\indent We have shown that a NN-MBE can be used to calculate the energy of methanol clusters with a speed up in the millions with respect to the MP2-MBE. The error of the NN-MBE is within m$E_h$s, which is  similar to the error of MP2-MBE with expansion up to three-body terms. The histogram of the errors of the NN-MBE display Gaussian shape, which makes the error per molecule decrease with the increase of system size. The satisfying accuracy and huge speed up enable the NN-MBE to treat large system with ab-inito accuracy, which would otherwise be impossible, such as treating solvation shell effects in ab-initio calculations. The Coulomb matrix is not invariant to permutations and even though we have shown that permutation invariance can be learned by augmenting the training samples with all of the possible permutations, it is still not perfectly invariant. Our current study focused on methanol and this scheme can easily be generalized to other systems. \\
\indent We introduced a new descriptor, the D-map, which is invertible with the geometry of a system. The D-map was able to predict the three-body energies reasonably well, and provides several advantages of its own. We then showed that a generative adversarial network could be trained on the D-map to provide hallucinated images, which are tunable and should be useful for inverse molecular design.

\section{Acknowledgement}
\indent We thank The University of Notre Dame's College of Science and Department of Chemistry and Biochemistry for generous start-up funding, and Nvidia corporation for some processors used in this work.



\bibliography{MBNN} 

\begin{thebibliography}{89}%
\makeatletter
\providecommand \@ifxundefined [1]{%
 \@ifx{#1\undefined}
}%
\providecommand \@ifnum [1]{%
 \ifnum #1\expandafter \@firstoftwo
 \else \expandafter \@secondoftwo
 \fi
}%
\providecommand \@ifx [1]{%
 \ifx #1\expandafter \@firstoftwo
 \else \expandafter \@secondoftwo
 \fi
}%
\providecommand \natexlab [1]{#1}%
\providecommand \enquote  [1]{``#1''}%
\providecommand \bibnamefont  [1]{#1}%
\providecommand \bibfnamefont [1]{#1}%
\providecommand \citenamefont [1]{#1}%
\providecommand \href@noop [0]{\@secondoftwo}%
\providecommand \href [0]{\begingroup \@sanitize@url \@href}%
\providecommand \@href[1]{\@@startlink{#1}\@@href}%
\providecommand \@@href[1]{\endgroup#1\@@endlink}%
\providecommand \@sanitize@url [0]{\catcode `\\12\catcode `\$12\catcode
  `\&12\catcode `\#12\catcode `\^12\catcode `\_12\catcode `\%12\relax}%
\providecommand \@@startlink[1]{}%
\providecommand \@@endlink[0]{}%
\providecommand \url  [0]{\begingroup\@sanitize@url \@url }%
\providecommand \@url [1]{\endgroup\@href {#1}{\urlprefix }}%
\providecommand \urlprefix  [0]{URL }%
\providecommand \Eprint [0]{\href }%
\providecommand \doibase [0]{http://dx.doi.org/}%
\providecommand \selectlanguage [0]{\@gobble}%
\providecommand \bibinfo  [0]{\@secondoftwo}%
\providecommand \bibfield  [0]{\@secondoftwo}%
\providecommand \translation [1]{[#1]}%
\providecommand \BibitemOpen [0]{}%
\providecommand \bibitemStop [0]{}%
\providecommand \bibitemNoStop [0]{.\EOS\space}%
\providecommand \EOS [0]{\spacefactor3000\relax}%
\providecommand \BibitemShut  [1]{\csname bibitem#1\endcsname}%
\let\auto@bib@innerbib\@empty
\bibitem [{\citenamefont {Raghavachari}\ and\ \citenamefont
  {Saha}(2015)}]{raghavachari2015accurate}%
  \BibitemOpen
  \bibfield  {author} {\bibinfo {author} {\bibfnamefont {K.}~\bibnamefont
  {Raghavachari}}\ and\ \bibinfo {author} {\bibfnamefont {A.}~\bibnamefont
  {Saha}},\ }\bibfield  {title} {\enquote {\bibinfo {title} {Accurate composite
  and fragment-based quantum chemical models for large molecules},}\
  }\href@noop {} {\bibfield  {journal} {\bibinfo  {journal} {Chem. Rev.}\
  }\textbf {\bibinfo {volume} {115}},\ \bibinfo {pages} {5643--5677} (\bibinfo
  {year} {2015})}\BibitemShut {NoStop}%
\bibitem [{\citenamefont {Mayhall}\ and\ \citenamefont
  {Raghavachari}(2012)}]{mayhall2012many}%
  \BibitemOpen
  \bibfield  {author} {\bibinfo {author} {\bibfnamefont {N.~J.}\ \bibnamefont
  {Mayhall}}\ and\ \bibinfo {author} {\bibfnamefont {K.}~\bibnamefont
  {Raghavachari}},\ }\bibfield  {title} {\enquote {\bibinfo {title}
  {Many-overlapping-body (mob) expansion: A generalized many body expansion for
  nondisjoint monomers in molecular fragmentation calculations of covalent
  molecules},}\ }\href@noop {} {\bibfield  {journal} {\bibinfo  {journal} {J.
  Chem. Theory Comput.}\ }\textbf {\bibinfo {volume} {8}},\ \bibinfo {pages}
  {2669--2675} (\bibinfo {year} {2012})}\BibitemShut {NoStop}%
\bibitem [{\citenamefont {Richard}\ and\ \citenamefont
  {Herbert}(2012)}]{richard2012generalized}%
  \BibitemOpen
  \bibfield  {author} {\bibinfo {author} {\bibfnamefont {R.~M.}\ \bibnamefont
  {Richard}}\ and\ \bibinfo {author} {\bibfnamefont {J.~M.}\ \bibnamefont
  {Herbert}},\ }\bibfield  {title} {\enquote {\bibinfo {title} {A generalized
  many-body expansion and a unified view of fragment-based methods in
  electronic structure theory},}\ }\href@noop {} {\bibfield  {journal}
  {\bibinfo  {journal} {J. Chem. Phys.}\ }\textbf {\bibinfo {volume} {137}},\
  \bibinfo {pages} {064113} (\bibinfo {year} {2012})}\BibitemShut {NoStop}%
\bibitem [{\citenamefont {Liu}\ and\ \citenamefont
  {Herbert}(2016)}]{liu2016pair}%
  \BibitemOpen
  \bibfield  {author} {\bibinfo {author} {\bibfnamefont {J.}~\bibnamefont
  {Liu}}\ and\ \bibinfo {author} {\bibfnamefont {J.~M.}\ \bibnamefont
  {Herbert}},\ }\bibfield  {title} {\enquote {\bibinfo {title} {Pair-pair
  approximation to the generalized many-body expansion: An alternative to the
  four-body expansion for ab initio prediction of protein energetics via
  molecular fragmentation},}\ }\href@noop {} {\bibfield  {journal} {\bibinfo
  {journal} {J. Chem. Theory Comput.}\ } (\bibinfo {year} {2016})}\BibitemShut
  {NoStop}%
\bibitem [{\citenamefont {Wen}\ \emph {et~al.}(2012)\citenamefont {Wen},
  \citenamefont {Nanda}, \citenamefont {Huang},\ and\ \citenamefont
  {Beran}}]{wen2012practical}%
  \BibitemOpen
  \bibfield  {author} {\bibinfo {author} {\bibfnamefont {S.}~\bibnamefont
  {Wen}}, \bibinfo {author} {\bibfnamefont {K.}~\bibnamefont {Nanda}}, \bibinfo
  {author} {\bibfnamefont {Y.}~\bibnamefont {Huang}}, \ and\ \bibinfo {author}
  {\bibfnamefont {G.~J.}\ \bibnamefont {Beran}},\ }\bibfield  {title} {\enquote
  {\bibinfo {title} {Practical quantum mechanics-based fragment methods for
  predicting molecular crystal properties},}\ }\href@noop {} {\bibfield
  {journal} {\bibinfo  {journal} {Phys. Chem. Chem. Phys.}\ }\textbf {\bibinfo
  {volume} {14}},\ \bibinfo {pages} {7578--7590} (\bibinfo {year}
  {2012})}\BibitemShut {NoStop}%
\bibitem [{\citenamefont {Beran}(2009)}]{beran2009approximating}%
  \BibitemOpen
  \bibfield  {author} {\bibinfo {author} {\bibfnamefont {G.~J.}\ \bibnamefont
  {Beran}},\ }\bibfield  {title} {\enquote {\bibinfo {title} {Approximating
  quantum many-body intermolecular interactions in molecular clusters using
  classical polarizable force fields},}\ }\href@noop {} {\bibfield  {journal}
  {\bibinfo  {journal} {J. Chem. Phys.}\ }\textbf {\bibinfo {volume} {130}},\
  \bibinfo {pages} {164115} (\bibinfo {year} {2009})}\BibitemShut {NoStop}%
\bibitem [{\citenamefont {Saha}\ and\ \citenamefont
  {Raghavachari}(2013)}]{saha2013dimers}%
  \BibitemOpen
  \bibfield  {author} {\bibinfo {author} {\bibfnamefont {A.}~\bibnamefont
  {Saha}}\ and\ \bibinfo {author} {\bibfnamefont {K.}~\bibnamefont
  {Raghavachari}},\ }\bibfield  {title} {\enquote {\bibinfo {title} {Dimers of
  dimers (dod): A new fragment-based method applied to large water clusters},}\
  }\href@noop {} {\bibfield  {journal} {\bibinfo  {journal} {J. Chem. Theory
  Comput.}\ }\textbf {\bibinfo {volume} {10}},\ \bibinfo {pages} {58--67}
  (\bibinfo {year} {2013})}\BibitemShut {NoStop}%
\bibitem [{\citenamefont {Gordon}\ \emph {et~al.}(2012)\citenamefont {Gordon},
  \citenamefont {Fedorov}, \citenamefont {Pruitt},\ and\ \citenamefont
  {Slipchenko}}]{gordon2012fragmentation}%
  \BibitemOpen
  \bibfield  {author} {\bibinfo {author} {\bibfnamefont {M.~S.}\ \bibnamefont
  {Gordon}}, \bibinfo {author} {\bibfnamefont {D.~G.}\ \bibnamefont {Fedorov}},
  \bibinfo {author} {\bibfnamefont {S.~R.}\ \bibnamefont {Pruitt}}, \ and\
  \bibinfo {author} {\bibfnamefont {L.~V.}\ \bibnamefont {Slipchenko}},\
  }\bibfield  {title} {\enquote {\bibinfo {title} {Fragmentation methods: a
  route to accurate calculations on large systems},}\ }\href@noop {} {\bibfield
   {journal} {\bibinfo  {journal} {Chem. Rev}\ }\textbf {\bibinfo {volume}
  {112}},\ \bibinfo {pages} {632--672} (\bibinfo {year} {2012})}\BibitemShut
  {NoStop}%
\bibitem [{\citenamefont {Dahlke}\ and\ \citenamefont
  {Truhlar}(2007{\natexlab{a}})}]{dahlke2007electrostatically}%
  \BibitemOpen
  \bibfield  {author} {\bibinfo {author} {\bibfnamefont {E.~E.}\ \bibnamefont
  {Dahlke}}\ and\ \bibinfo {author} {\bibfnamefont {D.~G.}\ \bibnamefont
  {Truhlar}},\ }\bibfield  {title} {\enquote {\bibinfo {title}
  {Electrostatically embedded many-body expansion for large systems, with
  applications to water clusters},}\ }\href@noop {} {\bibfield  {journal}
  {\bibinfo  {journal} {J. Chem. Theory Comput.}\ }\textbf {\bibinfo {volume}
  {3}},\ \bibinfo {pages} {46--53} (\bibinfo {year}
  {2007}{\natexlab{a}})}\BibitemShut {NoStop}%
\bibitem [{\citenamefont {Medders}, \citenamefont {Babin},\ and\ \citenamefont
  {Paesani}(2013)}]{medders2013critical}%
  \BibitemOpen
  \bibfield  {author} {\bibinfo {author} {\bibfnamefont {G.~R.}\ \bibnamefont
  {Medders}}, \bibinfo {author} {\bibfnamefont {V.}~\bibnamefont {Babin}}, \
  and\ \bibinfo {author} {\bibfnamefont {F.}~\bibnamefont {Paesani}},\
  }\bibfield  {title} {\enquote {\bibinfo {title} {A critical assessment of
  two-body and three-body interactions in water},}\ }\href@noop {} {\bibfield
  {journal} {\bibinfo  {journal} {J. Chem. Theory Comput.}\ }\textbf {\bibinfo
  {volume} {9}},\ \bibinfo {pages} {1103--1114} (\bibinfo {year}
  {2013})}\BibitemShut {NoStop}%
\bibitem [{\citenamefont {von Lilienfeld}\ and\ \citenamefont
  {Tkatchenko}(2010)}]{von2010two}%
  \BibitemOpen
  \bibfield  {author} {\bibinfo {author} {\bibfnamefont {O.~A.}\ \bibnamefont
  {von Lilienfeld}}\ and\ \bibinfo {author} {\bibfnamefont {A.}~\bibnamefont
  {Tkatchenko}},\ }\bibfield  {title} {\enquote {\bibinfo {title} {Two-and
  three-body interatomic dispersion energy contributions to binding in
  molecules and solids},}\ }\href@noop {} {\bibfield  {journal} {\bibinfo
  {journal} {The Journal of chemical physics}\ }\textbf {\bibinfo {volume}
  {132}},\ \bibinfo {pages} {234109} (\bibinfo {year} {2010})}\BibitemShut
  {NoStop}%
\bibitem [{\citenamefont {Jose}, \citenamefont {Beckett},\ and\ \citenamefont
  {Raghavachari}(2015)}]{jose2015vibrational}%
  \BibitemOpen
  \bibfield  {author} {\bibinfo {author} {\bibfnamefont {K.~J.}\ \bibnamefont
  {Jose}}, \bibinfo {author} {\bibfnamefont {D.}~\bibnamefont {Beckett}}, \
  and\ \bibinfo {author} {\bibfnamefont {K.}~\bibnamefont {Raghavachari}},\
  }\bibfield  {title} {\enquote {\bibinfo {title} {Vibrational circular
  dichroism spectra for large molecules through molecules-in-molecules
  fragment-based approach},}\ }\href@noop {} {\bibfield  {journal} {\bibinfo
  {journal} {J. Chem. Theory Comput.}\ }\textbf {\bibinfo {volume} {11}},\
  \bibinfo {pages} {4238--4247} (\bibinfo {year} {2015})}\BibitemShut {NoStop}%
\bibitem [{\citenamefont {Behler}\ and\ \citenamefont
  {Parrinello}(2007)}]{behler2007generalized}%
  \BibitemOpen
  \bibfield  {author} {\bibinfo {author} {\bibfnamefont {J.}~\bibnamefont
  {Behler}}\ and\ \bibinfo {author} {\bibfnamefont {M.}~\bibnamefont
  {Parrinello}},\ }\bibfield  {title} {\enquote {\bibinfo {title} {Generalized
  neural-network representation of high-dimensional potential-energy
  surfaces},}\ }\href@noop {} {\bibfield  {journal} {\bibinfo  {journal} {Phys.
  Rev. Lett.}\ }\textbf {\bibinfo {volume} {98}},\ \bibinfo {pages} {146401}
  (\bibinfo {year} {2007})}\BibitemShut {NoStop}%
\bibitem [{\citenamefont {Behler}(2011{\natexlab{a}})}]{behler2011neural}%
  \BibitemOpen
  \bibfield  {author} {\bibinfo {author} {\bibfnamefont {J.}~\bibnamefont
  {Behler}},\ }\bibfield  {title} {\enquote {\bibinfo {title} {Neural network
  potential-energy surfaces in chemistry: a tool for large-scale
  simulations},}\ }\href@noop {} {\bibfield  {journal} {\bibinfo  {journal}
  {Phys. Chem. Chem. Phys.}\ }\textbf {\bibinfo {volume} {13}},\ \bibinfo
  {pages} {17930--17955} (\bibinfo {year} {2011}{\natexlab{a}})}\BibitemShut
  {NoStop}%
\bibitem [{\citenamefont {Jiang}\ and\ \citenamefont
  {Guo}(2013)}]{jiang2013permutation}%
  \BibitemOpen
  \bibfield  {author} {\bibinfo {author} {\bibfnamefont {B.}~\bibnamefont
  {Jiang}}\ and\ \bibinfo {author} {\bibfnamefont {H.}~\bibnamefont {Guo}},\
  }\bibfield  {title} {\enquote {\bibinfo {title} {Permutation invariant
  polynomial neural network approach to fitting potential energy surfaces},}\
  }\href@noop {} {\bibfield  {journal} {\bibinfo  {journal} {J. Chem. Phys.}\
  }\textbf {\bibinfo {volume} {139}},\ \bibinfo {pages} {054112} (\bibinfo
  {year} {2013})}\BibitemShut {NoStop}%
\bibitem [{\citenamefont {Handley}\ and\ \citenamefont
  {Popelier}(2010)}]{handley2010potential}%
  \BibitemOpen
  \bibfield  {author} {\bibinfo {author} {\bibfnamefont {C.~M.}\ \bibnamefont
  {Handley}}\ and\ \bibinfo {author} {\bibfnamefont {P.~L.}\ \bibnamefont
  {Popelier}},\ }\bibfield  {title} {\enquote {\bibinfo {title} {Potential
  energy surfaces fitted by artificial neural networks},}\ }\href@noop {}
  {\bibfield  {journal} {\bibinfo  {journal} {J. Phys. Chem. A}\ }\textbf
  {\bibinfo {volume} {114}},\ \bibinfo {pages} {3371--3383} (\bibinfo {year}
  {2010})}\BibitemShut {NoStop}%
\bibitem [{\citenamefont {Cuny}\ \emph {et~al.}(2016)\citenamefont {Cuny},
  \citenamefont {Xie}, \citenamefont {Pickard},\ and\ \citenamefont
  {Hassanali}}]{cuny2016ab}%
  \BibitemOpen
  \bibfield  {author} {\bibinfo {author} {\bibfnamefont {J.}~\bibnamefont
  {Cuny}}, \bibinfo {author} {\bibfnamefont {Y.}~\bibnamefont {Xie}}, \bibinfo
  {author} {\bibfnamefont {C.~J.}\ \bibnamefont {Pickard}}, \ and\ \bibinfo
  {author} {\bibfnamefont {A.~A.}\ \bibnamefont {Hassanali}},\ }\bibfield
  {title} {\enquote {\bibinfo {title} {Ab initio quality nmr parameters in
  solid-state materials using a high-dimensional neural-network
  representation},}\ }\href@noop {} {\bibfield  {journal} {\bibinfo  {journal}
  {J. Chem. Theory Comput.}\ }\textbf {\bibinfo {volume} {12}},\ \bibinfo
  {pages} {765--773} (\bibinfo {year} {2016})}\BibitemShut {NoStop}%
\bibitem [{\citenamefont {Zhang}\ and\ \citenamefont
  {Zhang}(2014)}]{zhang2014effects}%
  \BibitemOpen
  \bibfield  {author} {\bibinfo {author} {\bibfnamefont {Z.}~\bibnamefont
  {Zhang}}\ and\ \bibinfo {author} {\bibfnamefont {D.~H.}\ \bibnamefont
  {Zhang}},\ }\bibfield  {title} {\enquote {\bibinfo {title} {Effects of
  reagent rotational excitation on the h+ chd3→ h2+ cd3 reaction: A seven
  dimensional time-dependent wave packet study},}\ }\href@noop {} {\bibfield
  {journal} {\bibinfo  {journal} {J. Chem. Phys.}\ }\textbf {\bibinfo {volume}
  {141}},\ \bibinfo {pages} {144309} (\bibinfo {year} {2014})}\BibitemShut
  {NoStop}%
\bibitem [{\citenamefont {Koch}\ and\ \citenamefont
  {Zhang}(2014)}]{koch2014communication}%
  \BibitemOpen
  \bibfield  {author} {\bibinfo {author} {\bibfnamefont {W.}~\bibnamefont
  {Koch}}\ and\ \bibinfo {author} {\bibfnamefont {D.~H.}\ \bibnamefont
  {Zhang}},\ }\bibfield  {title} {\enquote {\bibinfo {title} {Communication:
  Separable potential energy surfaces from multiplicative artificial neural
  networks},}\ }\href@noop {} {\bibfield  {journal} {\bibinfo  {journal} {J.
  Chem. Phys.}\ }\textbf {\bibinfo {volume} {141}},\ \bibinfo {pages} {021101}
  (\bibinfo {year} {2014})}\BibitemShut {NoStop}%
\bibitem [{\citenamefont {Chen}, \citenamefont {Xu},\ and\ \citenamefont
  {Zhang}(2013)}]{chen2013communication}%
  \BibitemOpen
  \bibfield  {author} {\bibinfo {author} {\bibfnamefont {J.}~\bibnamefont
  {Chen}}, \bibinfo {author} {\bibfnamefont {X.}~\bibnamefont {Xu}}, \ and\
  \bibinfo {author} {\bibfnamefont {D.~H.}\ \bibnamefont {Zhang}},\ }\bibfield
  {title} {\enquote {\bibinfo {title} {Communication: An accurate global
  potential energy surface for the oh+ co→ h+ co2 reaction using neural
  networks},}\ }\href@noop {} {\bibfield  {journal} {\bibinfo  {journal} {J.
  Chem. Phys.}\ }\textbf {\bibinfo {volume} {138}},\ \bibinfo {pages} {221104}
  (\bibinfo {year} {2013})}\BibitemShut {NoStop}%
\bibitem [{\citenamefont {Pinski}\ and\ \citenamefont
  {Csányi}(2013)}]{pinski2013reactive}%
  \BibitemOpen
  \bibfield  {author} {\bibinfo {author} {\bibfnamefont {P.}~\bibnamefont
  {Pinski}}\ and\ \bibinfo {author} {\bibfnamefont {G.}~\bibnamefont
  {Csányi}},\ }\bibfield  {title} {\enquote {\bibinfo {title} {Reactive
  many-body expansion for a protonated water cluster},}\ }\href@noop {}
  {\bibfield  {journal} {\bibinfo  {journal} {J. Chem. Theory Comput.}\
  }\textbf {\bibinfo {volume} {10}},\ \bibinfo {pages} {68--75} (\bibinfo
  {year} {2013})}\BibitemShut {NoStop}%
\bibitem [{\citenamefont {Xantheas}(1994)}]{xantheas1994abinitio}%
  \BibitemOpen
  \bibfield  {author} {\bibinfo {author} {\bibfnamefont {S.~S.}\ \bibnamefont
  {Xantheas}},\ }\bibfield  {title} {\enquote {\bibinfo {title} {Ab initio
  studies of cyclic water clusters (h2o) n, n = 1-6. ii. analysis of many-body
  interactions},}\ }\href@noop {} {\bibfield  {journal} {\bibinfo  {journal}
  {J. Chem. Phys.}\ }\textbf {\bibinfo {volume} {100}},\ \bibinfo {pages}
  {7523} (\bibinfo {year} {1994})}\BibitemShut {NoStop}%
\bibitem [{\citenamefont {Xantheas}(2000)}]{xantheas2000cooperativity}%
  \BibitemOpen
  \bibfield  {author} {\bibinfo {author} {\bibfnamefont {S.~S.}\ \bibnamefont
  {Xantheas}},\ }\bibfield  {title} {\enquote {\bibinfo {title} {Cooperativity
  and hydrogen bonding network in water clusters},}\ }\href@noop {} {\bibfield
  {journal} {\bibinfo  {journal} {Chem. Phys.}\ }\textbf {\bibinfo {volume}
  {258}},\ \bibinfo {pages} {225--231} (\bibinfo {year} {2000})}\BibitemShut
  {NoStop}%
\bibitem [{\citenamefont {Kulkarni}, \citenamefont {Ganesh},\ and\
  \citenamefont {Gadre}(2004)}]{kulkarni2004manybody}%
  \BibitemOpen
  \bibfield  {author} {\bibinfo {author} {\bibfnamefont {A.~D.}\ \bibnamefont
  {Kulkarni}}, \bibinfo {author} {\bibfnamefont {V.}~\bibnamefont {Ganesh}}, \
  and\ \bibinfo {author} {\bibfnamefont {S.~R.}\ \bibnamefont {Gadre}},\
  }\bibfield  {title} {\enquote {\bibinfo {title} {Many-body interaction
  analysis: Algorithm development and application to large molecular
  clusters},}\ }\href@noop {} {\bibfield  {journal} {\bibinfo  {journal} {J.
  Chem. Phys.}\ }\textbf {\bibinfo {volume} {121}},\ \bibinfo {pages} {5043}
  (\bibinfo {year} {2004})}\BibitemShut {NoStop}%
\bibitem [{\citenamefont {Ouyang}, \citenamefont {Cvitkovic},\ and\
  \citenamefont {Bettens}(2014)}]{ouyang2014trouble}%
  \BibitemOpen
  \bibfield  {author} {\bibinfo {author} {\bibfnamefont {J.~F.}\ \bibnamefont
  {Ouyang}}, \bibinfo {author} {\bibfnamefont {M.~W.}\ \bibnamefont
  {Cvitkovic}}, \ and\ \bibinfo {author} {\bibfnamefont {R.~P.}\ \bibnamefont
  {Bettens}},\ }\bibfield  {title} {\enquote {\bibinfo {title} {Trouble with
  the many-body expansion},}\ }\href@noop {} {\bibfield  {journal} {\bibinfo
  {journal} {J. Chem. Theory Comput.}\ }\textbf {\bibinfo {volume} {10}},\
  \bibinfo {pages} {3699--3707} (\bibinfo {year} {2014})}\BibitemShut {NoStop}%
\bibitem [{\citenamefont {Boys}\ and\ \citenamefont
  {Bernardi}(1970)}]{boys1970calculation}%
  \BibitemOpen
  \bibfield  {author} {\bibinfo {author} {\bibfnamefont {S.}~\bibnamefont
  {Boys}}\ and\ \bibinfo {author} {\bibfnamefont {F.}~\bibnamefont
  {Bernardi}},\ }\bibfield  {title} {\enquote {\bibinfo {title} {The
  calculation of small molecular interactions by the differences of separate
  total energies. some procedures with reduced errors},}\ }\href@noop {}
  {\bibfield  {journal} {\bibinfo  {journal} {Mol. Phys.}\ }\textbf {\bibinfo
  {volume} {19}},\ \bibinfo {pages} {553--566} (\bibinfo {year}
  {1970})}\BibitemShut {NoStop}%
\bibitem [{\citenamefont {Dahlke}\ and\ \citenamefont
  {Truhlar}(2007{\natexlab{b}})}]{dahlke2007electrostatically2}%
  \BibitemOpen
  \bibfield  {author} {\bibinfo {author} {\bibfnamefont {E.~E.}\ \bibnamefont
  {Dahlke}}\ and\ \bibinfo {author} {\bibfnamefont {D.~G.}\ \bibnamefont
  {Truhlar}},\ }\bibfield  {title} {\enquote {\bibinfo {title}
  {Electrostatically embedded many-body correlation energy, with applications
  to the calculation of accurate second-order m{\o}ller-plesset perturbation
  theory energies for large water clusters},}\ }\href@noop {} {\bibfield
  {journal} {\bibinfo  {journal} {J. Chem. Theory Comput.}\ }\textbf {\bibinfo
  {volume} {3}},\ \bibinfo {pages} {1342--1348} (\bibinfo {year}
  {2007}{\natexlab{b}})}\BibitemShut {NoStop}%
\bibitem [{\citenamefont {Richard}\ and\ \citenamefont
  {Herbert}(2013)}]{richard2013many-body}%
  \BibitemOpen
  \bibfield  {author} {\bibinfo {author} {\bibfnamefont {R.~M.}\ \bibnamefont
  {Richard}}\ and\ \bibinfo {author} {\bibfnamefont {J.~M.}\ \bibnamefont
  {Herbert}},\ }\bibfield  {title} {\enquote {\bibinfo {title} {Many-body
  expansion with overlapping fragments: Analysis of two approaches},}\ }\href
  {\doibase 10.1021/ct300985h} {\bibfield  {journal} {\bibinfo  {journal} {J.
  Chem. Theory Comput.}\ }\textbf {\bibinfo {volume} {9}},\ \bibinfo {pages}
  {1408--1416} (\bibinfo {year} {2013})},\ \bibinfo {note} {pMID: 26587603},\
  \Eprint {http://arxiv.org/abs/http://dx.doi.org/10.1021/ct300985h}
  {http://dx.doi.org/10.1021/ct300985h} \BibitemShut {NoStop}%
\bibitem [{\citenamefont {Lao}\ \emph {et~al.}(2016)\citenamefont {Lao},
  \citenamefont {Liu}, \citenamefont {Richard},\ and\ \citenamefont
  {Herbert}}]{lao2016understanding}%
  \BibitemOpen
  \bibfield  {author} {\bibinfo {author} {\bibfnamefont {K.~U.}\ \bibnamefont
  {Lao}}, \bibinfo {author} {\bibfnamefont {K.-Y.}\ \bibnamefont {Liu}},
  \bibinfo {author} {\bibfnamefont {R.~M.}\ \bibnamefont {Richard}}, \ and\
  \bibinfo {author} {\bibfnamefont {J.~M.}\ \bibnamefont {Herbert}},\
  }\bibfield  {title} {\enquote {\bibinfo {title} {Understanding the many-body
  expansion for large systems. ii. accuracy considerations},}\ }\href@noop {}
  {\bibfield  {journal} {\bibinfo  {journal} {J. Chem. Phys.}\ }\textbf
  {\bibinfo {volume} {144}},\ \bibinfo {pages} {164105} (\bibinfo {year}
  {2016})}\BibitemShut {NoStop}%
\bibitem [{\citenamefont
  {Behler}(2011{\natexlab{b}})}]{behler2011atomcentered}%
  \BibitemOpen
  \bibfield  {author} {\bibinfo {author} {\bibfnamefont {J.}~\bibnamefont
  {Behler}},\ }\bibfield  {title} {\enquote {\bibinfo {title} {Atom-centered
  symmetry functions for constructing high-dimensional neural network
  potentials},}\ }\href@noop {} {\bibfield  {journal} {\bibinfo  {journal} {J.
  Chem. Phys.}\ }\textbf {\bibinfo {volume} {134}},\ \bibinfo {pages} {074106}
  (\bibinfo {year} {2011}{\natexlab{b}})}\BibitemShut {NoStop}%
\bibitem [{\citenamefont {Khaliullin}\ \emph {et~al.}(2011)\citenamefont
  {Khaliullin}, \citenamefont {Eshet}, \citenamefont {K{\"u}hne}, \citenamefont
  {Behler},\ and\ \citenamefont {Parrinello}}]{khaliullin2011nucleation}%
  \BibitemOpen
  \bibfield  {author} {\bibinfo {author} {\bibfnamefont {R.~Z.}\ \bibnamefont
  {Khaliullin}}, \bibinfo {author} {\bibfnamefont {H.}~\bibnamefont {Eshet}},
  \bibinfo {author} {\bibfnamefont {T.~D.}\ \bibnamefont {K{\"u}hne}}, \bibinfo
  {author} {\bibfnamefont {J.}~\bibnamefont {Behler}}, \ and\ \bibinfo {author}
  {\bibfnamefont {M.}~\bibnamefont {Parrinello}},\ }\bibfield  {title}
  {\enquote {\bibinfo {title} {Nucleation mechanism for the direct
  graphite-to-diamond phase transition},}\ }\href@noop {} {\bibfield  {journal}
  {\bibinfo  {journal} {Nat. Mater.}\ }\textbf {\bibinfo {volume} {10}},\
  \bibinfo {pages} {693--697} (\bibinfo {year} {2011})}\BibitemShut {NoStop}%
\bibitem [{\citenamefont {Xin}, \citenamefont {Chen},\ and\ \citenamefont
  {Zhang}(2014)}]{xu2014global}%
  \BibitemOpen
  \bibfield  {author} {\bibinfo {author} {\bibfnamefont {X.}~\bibnamefont
  {Xin}}, \bibinfo {author} {\bibfnamefont {J.}~\bibnamefont {Chen}}, \ and\
  \bibinfo {author} {\bibfnamefont {D.~H.}\ \bibnamefont {Zhang}},\ }\bibfield
  {title} {\enquote {\bibinfo {title} {Global potential energy surface for the
  h+ch$4$<->h2+ch3 reaction using neural networks},}\ }\href@noop {} {\bibfield
   {journal} {\bibinfo  {journal} {Chin. J. Chem. Phys.}\ }\textbf {\bibinfo
  {volume} {27}},\ \bibinfo {pages} {373} (\bibinfo {year} {2014})}\BibitemShut
  {NoStop}%
\bibitem [{\citenamefont {Shao}\ \emph {et~al.}(2016)\citenamefont {Shao},
  \citenamefont {Chen}, \citenamefont {Zhao},\ and\ \citenamefont
  {Zhang}}]{shao2016communication}%
  \BibitemOpen
  \bibfield  {author} {\bibinfo {author} {\bibfnamefont {K.}~\bibnamefont
  {Shao}}, \bibinfo {author} {\bibfnamefont {J.}~\bibnamefont {Chen}}, \bibinfo
  {author} {\bibfnamefont {Z.}~\bibnamefont {Zhao}}, \ and\ \bibinfo {author}
  {\bibfnamefont {D.~H.}\ \bibnamefont {Zhang}},\ }\bibfield  {title} {\enquote
  {\bibinfo {title} {Communication: Fitting potential energy surfaces with
  fundamental invariant neural network},}\ }\href@noop {} {\bibfield  {journal}
  {\bibinfo  {journal} {J. Chem. Phys.}\ }\textbf {\bibinfo {volume} {145}},\
  \bibinfo {pages} {071101} (\bibinfo {year} {2016})}\BibitemShut {NoStop}%
\bibitem [{\citenamefont {Li}\ and\ \citenamefont
  {Guo}(2015)}]{li2015permutationally}%
  \BibitemOpen
  \bibfield  {author} {\bibinfo {author} {\bibfnamefont {J.}~\bibnamefont
  {Li}}\ and\ \bibinfo {author} {\bibfnamefont {H.}~\bibnamefont {Guo}},\
  }\bibfield  {title} {\enquote {\bibinfo {title} {Permutationally invariant
  fitting of intermolecular potential energy surfaces: A case study of the
  ne-c2h2 system},}\ }\href@noop {} {\bibfield  {journal} {\bibinfo  {journal}
  {J. Chem. Phys.}\ }\textbf {\bibinfo {volume} {143}},\ \bibinfo {pages}
  {214304} (\bibinfo {year} {2015})}\BibitemShut {NoStop}%
\bibitem [{\citenamefont {Li}\ \emph {et~al.}(2015)\citenamefont {Li},
  \citenamefont {Chen}, \citenamefont {Zhao}, \citenamefont {Xie},
  \citenamefont {Zhang},\ and\ \citenamefont {Guo}}]{li2015permutationally2}%
  \BibitemOpen
  \bibfield  {author} {\bibinfo {author} {\bibfnamefont {J.}~\bibnamefont
  {Li}}, \bibinfo {author} {\bibfnamefont {J.}~\bibnamefont {Chen}}, \bibinfo
  {author} {\bibfnamefont {Z.}~\bibnamefont {Zhao}}, \bibinfo {author}
  {\bibfnamefont {D.}~\bibnamefont {Xie}}, \bibinfo {author} {\bibfnamefont
  {D.~H.}\ \bibnamefont {Zhang}}, \ and\ \bibinfo {author} {\bibfnamefont
  {H.}~\bibnamefont {Guo}},\ }\bibfield  {title} {\enquote {\bibinfo {title} {A
  permutationally invariant full-dimensional ab initio potential energy surface
  for the abstraction and exchange channels of the h+ ch4 system},}\
  }\href@noop {} {\bibfield  {journal} {\bibinfo  {journal} {J. Chem. Phys.}\
  }\textbf {\bibinfo {volume} {142}},\ \bibinfo {pages} {204302} (\bibinfo
  {year} {2015})}\BibitemShut {NoStop}%
\bibitem [{\citenamefont {Medders}\ \emph {et~al.}(2015)\citenamefont
  {Medders}, \citenamefont {G{\"o}tz}, \citenamefont {Morales}, \citenamefont
  {Bajaj},\ and\ \citenamefont {Paesani}}]{medders2015representation}%
  \BibitemOpen
  \bibfield  {author} {\bibinfo {author} {\bibfnamefont {G.~R.}\ \bibnamefont
  {Medders}}, \bibinfo {author} {\bibfnamefont {A.~W.}\ \bibnamefont
  {G{\"o}tz}}, \bibinfo {author} {\bibfnamefont {M.~A.}\ \bibnamefont
  {Morales}}, \bibinfo {author} {\bibfnamefont {P.}~\bibnamefont {Bajaj}}, \
  and\ \bibinfo {author} {\bibfnamefont {F.}~\bibnamefont {Paesani}},\
  }\bibfield  {title} {\enquote {\bibinfo {title} {On the representation of
  many-body interactions in water},}\ }\href@noop {} {\bibfield  {journal}
  {\bibinfo  {journal} {J. Chem. Phys.}\ }\textbf {\bibinfo {volume} {143}},\
  \bibinfo {pages} {104102} (\bibinfo {year} {2015})}\BibitemShut {NoStop}%
\bibitem [{\citenamefont {Conte}, \citenamefont {Qu},\ and\ \citenamefont
  {Bowman}(2015)}]{conte2015permutationally}%
  \BibitemOpen
  \bibfield  {author} {\bibinfo {author} {\bibfnamefont {R.}~\bibnamefont
  {Conte}}, \bibinfo {author} {\bibfnamefont {C.}~\bibnamefont {Qu}}, \ and\
  \bibinfo {author} {\bibfnamefont {J.~M.}\ \bibnamefont {Bowman}},\ }\bibfield
   {title} {\enquote {\bibinfo {title} {Permutationally invariant fitting of
  many-body, non-covalent interactions with application to three-body
  methane--water--water},}\ }\href@noop {} {\bibfield  {journal} {\bibinfo
  {journal} {J. Chem. Theory Comput.}\ }\textbf {\bibinfo {volume} {11}},\
  \bibinfo {pages} {1631--1638} (\bibinfo {year} {2015})}\BibitemShut {NoStop}%
\bibitem [{\citenamefont {Rupp}(2015)}]{Rupp:2015aa}%
  \BibitemOpen
  \bibfield  {author} {\bibinfo {author} {\bibfnamefont {M.}~\bibnamefont
  {Rupp}},\ }\bibfield  {title} {\enquote {\bibinfo {title} {Machine learning
  for quantum mechanics in a nutshell},}\ }\href {\doibase 10.1002/qua.24954}
  {\bibfield  {journal} {\bibinfo  {journal} {Int. J. Quantum Chem.}\ }\textbf
  {\bibinfo {volume} {115}},\ \bibinfo {pages} {1058--1073} (\bibinfo {year}
  {2015})}\BibitemShut {NoStop}%
\bibitem [{\citenamefont {Hansen}\ \emph {et~al.}(2015)\citenamefont {Hansen},
  \citenamefont {Biegler}, \citenamefont {Ramakrishnan}, \citenamefont
  {Pronobis}, \citenamefont {von Lilienfeld}, \citenamefont {M{\"u}ller},\ and\
  \citenamefont {Tkatchenko}}]{Hansen:2015aa}%
  \BibitemOpen
  \bibfield  {author} {\bibinfo {author} {\bibfnamefont {K.}~\bibnamefont
  {Hansen}}, \bibinfo {author} {\bibfnamefont {F.}~\bibnamefont {Biegler}},
  \bibinfo {author} {\bibfnamefont {R.}~\bibnamefont {Ramakrishnan}}, \bibinfo
  {author} {\bibfnamefont {W.}~\bibnamefont {Pronobis}}, \bibinfo {author}
  {\bibfnamefont {O.~A.}\ \bibnamefont {von Lilienfeld}}, \bibinfo {author}
  {\bibfnamefont {K.-R.}\ \bibnamefont {M{\"u}ller}}, \ and\ \bibinfo {author}
  {\bibfnamefont {A.}~\bibnamefont {Tkatchenko}},\ }\bibfield  {title}
  {\enquote {\bibinfo {title} {Machine learning predictions of molecular
  properties: Accurate many-body potentials and nonlocality in chemical
  space},}\ }\href {\doibase 10.1021/acs.jpclett.5b00831} {\bibfield  {journal}
  {\bibinfo  {journal} {J. Phys. Chem. Lett.}\ }\textbf {\bibinfo {volume}
  {6}},\ \bibinfo {pages} {2326--2331} (\bibinfo {year} {2015})},\ \Eprint
  {http://arxiv.org/abs/http://dx.doi.org/10.1021/acs.jpclett.5b00831}
  {http://dx.doi.org/10.1021/acs.jpclett.5b00831} \BibitemShut {NoStop}%
\bibitem [{\citenamefont {Montavon}\ \emph {et~al.}(2013)\citenamefont
  {Montavon}, \citenamefont {Rupp}, \citenamefont {Gobre}, \citenamefont
  {Vazquez-Mayagoitia}, \citenamefont {Hansen}, \citenamefont {Tkatchenko},
  \citenamefont {M{\"u}ller},\ and\ \citenamefont {von
  Lilienfeld}}]{montavon2013machine}%
  \BibitemOpen
  \bibfield  {author} {\bibinfo {author} {\bibfnamefont {G.}~\bibnamefont
  {Montavon}}, \bibinfo {author} {\bibfnamefont {M.}~\bibnamefont {Rupp}},
  \bibinfo {author} {\bibfnamefont {V.}~\bibnamefont {Gobre}}, \bibinfo
  {author} {\bibfnamefont {A.}~\bibnamefont {Vazquez-Mayagoitia}}, \bibinfo
  {author} {\bibfnamefont {K.}~\bibnamefont {Hansen}}, \bibinfo {author}
  {\bibfnamefont {A.}~\bibnamefont {Tkatchenko}}, \bibinfo {author}
  {\bibfnamefont {K.-R.}\ \bibnamefont {M{\"u}ller}}, \ and\ \bibinfo {author}
  {\bibfnamefont {O.~A.}\ \bibnamefont {von Lilienfeld}},\ }\bibfield  {title}
  {\enquote {\bibinfo {title} {Machine learning of molecular electronic
  properties in chemical compound space},}\ }\href@noop {} {\bibfield
  {journal} {\bibinfo  {journal} {New J. Phys.}\ }\textbf {\bibinfo {volume}
  {15}},\ \bibinfo {pages} {095003} (\bibinfo {year} {2013})}\BibitemShut
  {NoStop}%
\bibitem [{\citenamefont {Pilania}\ \emph {et~al.}(2013)\citenamefont
  {Pilania}, \citenamefont {Wang}, \citenamefont {Jiang}, \citenamefont
  {Rajasekaran},\ and\ \citenamefont {Ramprasad}}]{pilania2013accelerating}%
  \BibitemOpen
  \bibfield  {author} {\bibinfo {author} {\bibfnamefont {G.}~\bibnamefont
  {Pilania}}, \bibinfo {author} {\bibfnamefont {C.}~\bibnamefont {Wang}},
  \bibinfo {author} {\bibfnamefont {X.}~\bibnamefont {Jiang}}, \bibinfo
  {author} {\bibfnamefont {S.}~\bibnamefont {Rajasekaran}}, \ and\ \bibinfo
  {author} {\bibfnamefont {R.}~\bibnamefont {Ramprasad}},\ }\bibfield  {title}
  {\enquote {\bibinfo {title} {Accelerating materials property predictions
  using machine learning},}\ }\href@noop {} {\bibfield  {journal} {\bibinfo
  {journal} {Sci. Rep.}\ }\textbf {\bibinfo {volume} {3}} (\bibinfo {year}
  {2013})}\BibitemShut {NoStop}%
\bibitem [{\citenamefont {Ghasemi}\ \emph {et~al.}(2015)\citenamefont
  {Ghasemi}, \citenamefont {Hofstetter}, \citenamefont {Saha},\ and\
  \citenamefont {Goedecker}}]{ghasemi2015interatomic}%
  \BibitemOpen
  \bibfield  {author} {\bibinfo {author} {\bibfnamefont {S.~A.}\ \bibnamefont
  {Ghasemi}}, \bibinfo {author} {\bibfnamefont {A.}~\bibnamefont {Hofstetter}},
  \bibinfo {author} {\bibfnamefont {S.}~\bibnamefont {Saha}}, \ and\ \bibinfo
  {author} {\bibfnamefont {S.}~\bibnamefont {Goedecker}},\ }\bibfield  {title}
  {\enquote {\bibinfo {title} {Interatomic potentials for ionic systems with
  density functional accuracy based on charge densities obtained by a neural
  network},}\ }\href@noop {} {\bibfield  {journal} {\bibinfo  {journal} {Phys.
  Rev. B}\ }\textbf {\bibinfo {volume} {92}},\ \bibinfo {pages} {045131}
  (\bibinfo {year} {2015})}\BibitemShut {NoStop}%
\bibitem [{\citenamefont {Sch{\"u}tt}\ \emph {et~al.}(2014)\citenamefont
  {Sch{\"u}tt}, \citenamefont {Glawe}, \citenamefont {Brockherde},
  \citenamefont {Sanna}, \citenamefont {M{\"u}ller},\ and\ \citenamefont
  {Gross}}]{schutt2014represent}%
  \BibitemOpen
  \bibfield  {author} {\bibinfo {author} {\bibfnamefont {K.}~\bibnamefont
  {Sch{\"u}tt}}, \bibinfo {author} {\bibfnamefont {H.}~\bibnamefont {Glawe}},
  \bibinfo {author} {\bibfnamefont {F.}~\bibnamefont {Brockherde}}, \bibinfo
  {author} {\bibfnamefont {A.}~\bibnamefont {Sanna}}, \bibinfo {author}
  {\bibfnamefont {K.}~\bibnamefont {M{\"u}ller}}, \ and\ \bibinfo {author}
  {\bibfnamefont {E.}~\bibnamefont {Gross}},\ }\bibfield  {title} {\enquote
  {\bibinfo {title} {How to represent crystal structures for machine learning:
  Towards fast prediction of electronic properties},}\ }\href@noop {}
  {\bibfield  {journal} {\bibinfo  {journal} {Phys. Rev. B}\ }\textbf {\bibinfo
  {volume} {89}},\ \bibinfo {pages} {205118} (\bibinfo {year}
  {2014})}\BibitemShut {NoStop}%
\bibitem [{\citenamefont {Olivares-Amaya}\ \emph {et~al.}(2011)\citenamefont
  {Olivares-Amaya}, \citenamefont {Amador-Bedolla}, \citenamefont {Hachmann},
  \citenamefont {Atahan-Evrenk}, \citenamefont {S{\'a}nchez-Carrera},
  \citenamefont {Vogt},\ and\ \citenamefont
  {Aspuru-Guzik}}]{olivares2011accelerated}%
  \BibitemOpen
  \bibfield  {author} {\bibinfo {author} {\bibfnamefont {R.}~\bibnamefont
  {Olivares-Amaya}}, \bibinfo {author} {\bibfnamefont {C.}~\bibnamefont
  {Amador-Bedolla}}, \bibinfo {author} {\bibfnamefont {J.}~\bibnamefont
  {Hachmann}}, \bibinfo {author} {\bibfnamefont {S.}~\bibnamefont
  {Atahan-Evrenk}}, \bibinfo {author} {\bibfnamefont {R.~S.}\ \bibnamefont
  {S{\'a}nchez-Carrera}}, \bibinfo {author} {\bibfnamefont {L.}~\bibnamefont
  {Vogt}}, \ and\ \bibinfo {author} {\bibfnamefont {A.}~\bibnamefont
  {Aspuru-Guzik}},\ }\bibfield  {title} {\enquote {\bibinfo {title}
  {Accelerated computational discovery of high-performance materials for
  organic photovoltaics by means of cheminformatics},}\ }\href@noop {}
  {\bibfield  {journal} {\bibinfo  {journal} {Energ. Environ. Sci.}\ }\textbf
  {\bibinfo {volume} {4}},\ \bibinfo {pages} {4849--4861} (\bibinfo {year}
  {2011})}\BibitemShut {NoStop}%
\bibitem [{\citenamefont {Ma}\ \emph {et~al.}(2015)\citenamefont {Ma},
  \citenamefont {Li}, \citenamefont {Achenie},\ and\ \citenamefont
  {Xin}}]{ma2015machine}%
  \BibitemOpen
  \bibfield  {author} {\bibinfo {author} {\bibfnamefont {X.}~\bibnamefont
  {Ma}}, \bibinfo {author} {\bibfnamefont {Z.}~\bibnamefont {Li}}, \bibinfo
  {author} {\bibfnamefont {L.~E.}\ \bibnamefont {Achenie}}, \ and\ \bibinfo
  {author} {\bibfnamefont {H.}~\bibnamefont {Xin}},\ }\bibfield  {title}
  {\enquote {\bibinfo {title} {Machine-learning-augmented chemisorption model
  for co2 electroreduction catalyst screening},}\ }\href@noop {} {\bibfield
  {journal} {\bibinfo  {journal} {J. Phys. Chem. Lett.}\ }\textbf {\bibinfo
  {volume} {6}},\ \bibinfo {pages} {3528--3533} (\bibinfo {year}
  {2015})}\BibitemShut {NoStop}%
\bibitem [{\citenamefont {Ediz}\ \emph {et~al.}(2009)\citenamefont {Ediz},
  \citenamefont {Monda}, \citenamefont {Brown},\ and\ \citenamefont
  {Yaron}}]{ediz2009using}%
  \BibitemOpen
  \bibfield  {author} {\bibinfo {author} {\bibfnamefont {V.}~\bibnamefont
  {Ediz}}, \bibinfo {author} {\bibfnamefont {A.~C.}\ \bibnamefont {Monda}},
  \bibinfo {author} {\bibfnamefont {R.~P.}\ \bibnamefont {Brown}}, \ and\
  \bibinfo {author} {\bibfnamefont {D.~J.}\ \bibnamefont {Yaron}},\ }\bibfield
  {title} {\enquote {\bibinfo {title} {Using molecular similarity to develop
  reliable models of chemical reactions in complex environments},}\ }\href@noop
  {} {\bibfield  {journal} {\bibinfo  {journal} {Journal of chemical theory and
  computation}\ }\textbf {\bibinfo {volume} {5}},\ \bibinfo {pages}
  {3175--3184} (\bibinfo {year} {2009})}\BibitemShut {NoStop}%
\bibitem [{\citenamefont {Lopez-Bezanilla}\ and\ \citenamefont {von
  Lilienfeld}(2014)}]{lopez2014modeling}%
  \BibitemOpen
  \bibfield  {author} {\bibinfo {author} {\bibfnamefont {A.}~\bibnamefont
  {Lopez-Bezanilla}}\ and\ \bibinfo {author} {\bibfnamefont {O.~A.}\
  \bibnamefont {von Lilienfeld}},\ }\bibfield  {title} {\enquote {\bibinfo
  {title} {Modeling electronic quantum transport with machine learning},}\
  }\href@noop {} {\bibfield  {journal} {\bibinfo  {journal} {Phys. Rev. B}\
  }\textbf {\bibinfo {volume} {89}},\ \bibinfo {pages} {235411} (\bibinfo
  {year} {2014})}\BibitemShut {NoStop}%
\bibitem [{\citenamefont {Snyder}\ \emph {et~al.}(2013)\citenamefont {Snyder},
  \citenamefont {Rupp}, \citenamefont {Hansen}, \citenamefont {Blooston},
  \citenamefont {M{\"u}ller},\ and\ \citenamefont {Burke}}]{Snyder:2013aa}%
  \BibitemOpen
  \bibfield  {author} {\bibinfo {author} {\bibfnamefont {J.~C.}\ \bibnamefont
  {Snyder}}, \bibinfo {author} {\bibfnamefont {M.}~\bibnamefont {Rupp}},
  \bibinfo {author} {\bibfnamefont {K.}~\bibnamefont {Hansen}}, \bibinfo
  {author} {\bibfnamefont {L.}~\bibnamefont {Blooston}}, \bibinfo {author}
  {\bibfnamefont {K.-R.}\ \bibnamefont {M{\"u}ller}}, \ and\ \bibinfo {author}
  {\bibfnamefont {K.}~\bibnamefont {Burke}},\ }\bibfield  {title} {\enquote
  {\bibinfo {title} {Orbital-free bond breaking via machine learning},}\ }\href
  {\doibase http://dx.doi.org/10.1063/1.4834075} {\bibfield  {journal}
  {\bibinfo  {journal} {J. Chem. Phys.}\ }\textbf {\bibinfo {volume} {139}},\
  \bibinfo {pages} {224104} (\bibinfo {year} {2013})}\BibitemShut {NoStop}%
\bibitem [{\citenamefont {Snyder}\ \emph {et~al.}(2012)\citenamefont {Snyder},
  \citenamefont {Rupp}, \citenamefont {Hansen}, \citenamefont {M{\"u}ller},\
  and\ \citenamefont {Burke}}]{snyder2012finding}%
  \BibitemOpen
  \bibfield  {author} {\bibinfo {author} {\bibfnamefont {J.~C.}\ \bibnamefont
  {Snyder}}, \bibinfo {author} {\bibfnamefont {M.}~\bibnamefont {Rupp}},
  \bibinfo {author} {\bibfnamefont {K.}~\bibnamefont {Hansen}}, \bibinfo
  {author} {\bibfnamefont {K.-R.}\ \bibnamefont {M{\"u}ller}}, \ and\ \bibinfo
  {author} {\bibfnamefont {K.}~\bibnamefont {Burke}},\ }\bibfield  {title}
  {\enquote {\bibinfo {title} {Finding density functionals with machine
  learning},}\ }\href@noop {} {\bibfield  {journal} {\bibinfo  {journal} {Phys.
  Rev. Lett.}\ }\textbf {\bibinfo {volume} {108}},\ \bibinfo {pages} {253002}
  (\bibinfo {year} {2012})}\BibitemShut {NoStop}%
\bibitem [{\citenamefont {Yao}\ and\ \citenamefont
  {Parkhill}(2016)}]{yao2016kinetic}%
  \BibitemOpen
  \bibfield  {author} {\bibinfo {author} {\bibfnamefont {K.}~\bibnamefont
  {Yao}}\ and\ \bibinfo {author} {\bibfnamefont {J.}~\bibnamefont {Parkhill}},\
  }\bibfield  {title} {\enquote {\bibinfo {title} {Kinetic energy of
  hydrocarbons as a function of electron density and convolutional neural
  networks},}\ }\href@noop {} {\bibfield  {journal} {\bibinfo  {journal} {J.
  Chem. Theory Comput.}\ }\textbf {\bibinfo {volume} {12}},\ \bibinfo {pages}
  {1139--1147} (\bibinfo {year} {2016})}\BibitemShut {NoStop}%
\bibitem [{\citenamefont {Hachmann}\ \emph {et~al.}(2011)\citenamefont
  {Hachmann}, \citenamefont {Olivares-Amaya}, \citenamefont {Atahan-Evrenk},
  \citenamefont {Amador-Bedolla}, \citenamefont {S{\'a}nchez-Carrera},
  \citenamefont {Gold-Parker}, \citenamefont {Vogt}, \citenamefont {Brockway},\
  and\ \citenamefont {Aspuru-Guzik}}]{hachmann2011harvard}%
  \BibitemOpen
  \bibfield  {author} {\bibinfo {author} {\bibfnamefont {J.}~\bibnamefont
  {Hachmann}}, \bibinfo {author} {\bibfnamefont {R.}~\bibnamefont
  {Olivares-Amaya}}, \bibinfo {author} {\bibfnamefont {S.}~\bibnamefont
  {Atahan-Evrenk}}, \bibinfo {author} {\bibfnamefont {C.}~\bibnamefont
  {Amador-Bedolla}}, \bibinfo {author} {\bibfnamefont {R.~S.}\ \bibnamefont
  {S{\'a}nchez-Carrera}}, \bibinfo {author} {\bibfnamefont {A.}~\bibnamefont
  {Gold-Parker}}, \bibinfo {author} {\bibfnamefont {L.}~\bibnamefont {Vogt}},
  \bibinfo {author} {\bibfnamefont {A.~M.}\ \bibnamefont {Brockway}}, \ and\
  \bibinfo {author} {\bibfnamefont {A.}~\bibnamefont {Aspuru-Guzik}},\
  }\bibfield  {title} {\enquote {\bibinfo {title} {The harvard clean energy
  project: large-scale computational screening and design of organic
  photovoltaics on the world community grid},}\ }\href@noop {} {\bibfield
  {journal} {\bibinfo  {journal} {J. Phys. Chem. Lett.}\ }\textbf {\bibinfo
  {volume} {2}},\ \bibinfo {pages} {2241--2251} (\bibinfo {year}
  {2011})}\BibitemShut {NoStop}%
\bibitem [{\citenamefont {Hachmann}\ \emph {et~al.}(2014)\citenamefont
  {Hachmann}, \citenamefont {Olivares-Amaya}, \citenamefont {Jinich},
  \citenamefont {Appleton}, \citenamefont {Blood-Forsythe}, \citenamefont
  {Seress}, \citenamefont {Roman-Salgado}, \citenamefont {Trepte},
  \citenamefont {Atahan-Evrenk},\ and\ \citenamefont {Er}}]{hachmann2014lead}%
  \BibitemOpen
  \bibfield  {author} {\bibinfo {author} {\bibfnamefont {J.}~\bibnamefont
  {Hachmann}}, \bibinfo {author} {\bibfnamefont {R.}~\bibnamefont
  {Olivares-Amaya}}, \bibinfo {author} {\bibfnamefont {A.}~\bibnamefont
  {Jinich}}, \bibinfo {author} {\bibfnamefont {A.~L.}\ \bibnamefont
  {Appleton}}, \bibinfo {author} {\bibfnamefont {M.~A.}\ \bibnamefont
  {Blood-Forsythe}}, \bibinfo {author} {\bibfnamefont {L.~R.}\ \bibnamefont
  {Seress}}, \bibinfo {author} {\bibfnamefont {C.}~\bibnamefont
  {Roman-Salgado}}, \bibinfo {author} {\bibfnamefont {K.}~\bibnamefont
  {Trepte}}, \bibinfo {author} {\bibfnamefont {S.}~\bibnamefont
  {Atahan-Evrenk}}, \ and\ \bibinfo {author} {\bibfnamefont {S.}~\bibnamefont
  {Er}},\ }\bibfield  {title} {\enquote {\bibinfo {title} {Lead candidates for
  high-performance organic photovoltaics from high-throughput quantum
  chemistry--the harvard clean energy project},}\ }\href@noop {} {\bibfield
  {journal} {\bibinfo  {journal} {Energ. Environ. Sci.}\ }\textbf {\bibinfo
  {volume} {7}},\ \bibinfo {pages} {698--704} (\bibinfo {year}
  {2014})}\BibitemShut {NoStop}%
\bibitem [{\citenamefont {Hautier}\ \emph {et~al.}(2010)\citenamefont
  {Hautier}, \citenamefont {Fischer}, \citenamefont {Jain}, \citenamefont
  {Mueller},\ and\ \citenamefont {Ceder}}]{hautier2010finding}%
  \BibitemOpen
  \bibfield  {author} {\bibinfo {author} {\bibfnamefont {G.}~\bibnamefont
  {Hautier}}, \bibinfo {author} {\bibfnamefont {C.~C.}\ \bibnamefont
  {Fischer}}, \bibinfo {author} {\bibfnamefont {A.}~\bibnamefont {Jain}},
  \bibinfo {author} {\bibfnamefont {T.}~\bibnamefont {Mueller}}, \ and\
  \bibinfo {author} {\bibfnamefont {G.}~\bibnamefont {Ceder}},\ }\bibfield
  {title} {\enquote {\bibinfo {title} {Finding nature’s missing ternary oxide
  compounds using machine learning and density functional theory},}\
  }\href@noop {} {\bibfield  {journal} {\bibinfo  {journal} {Chem. Mater.}\
  }\textbf {\bibinfo {volume} {22}},\ \bibinfo {pages} {3762--3767} (\bibinfo
  {year} {2010})}\BibitemShut {NoStop}%
\bibitem [{\citenamefont {Ediz}\ \emph {et~al.}(2008)\citenamefont {Ediz},
  \citenamefont {Lee}, \citenamefont {Armitage},\ and\ \citenamefont
  {Yaron}}]{ediz2008molecular}%
  \BibitemOpen
  \bibfield  {author} {\bibinfo {author} {\bibfnamefont {V.}~\bibnamefont
  {Ediz}}, \bibinfo {author} {\bibfnamefont {J.~L.}\ \bibnamefont {Lee}},
  \bibinfo {author} {\bibfnamefont {B.~A.}\ \bibnamefont {Armitage}}, \ and\
  \bibinfo {author} {\bibfnamefont {D.}~\bibnamefont {Yaron}},\ }\bibfield
  {title} {\enquote {\bibinfo {title} {Molecular engineering of torsional
  potentials in fluorogenic dyes via electronic substituent effects},}\
  }\href@noop {} {\bibfield  {journal} {\bibinfo  {journal} {The Journal of
  Physical Chemistry A}\ }\textbf {\bibinfo {volume} {112}},\ \bibinfo {pages}
  {9692--9701} (\bibinfo {year} {2008})}\BibitemShut {NoStop}%
\bibitem [{\citenamefont {Malshe}\ \emph {et~al.}(2009)\citenamefont {Malshe},
  \citenamefont {Narulkar}, \citenamefont {Raff}, \citenamefont {Hagan},
  \citenamefont {Bukkapatnam}, \citenamefont {Agrawal},\ and\ \citenamefont
  {Komanduri}}]{malshe2009development}%
  \BibitemOpen
  \bibfield  {author} {\bibinfo {author} {\bibfnamefont {M.}~\bibnamefont
  {Malshe}}, \bibinfo {author} {\bibfnamefont {R.}~\bibnamefont {Narulkar}},
  \bibinfo {author} {\bibfnamefont {L.}~\bibnamefont {Raff}}, \bibinfo {author}
  {\bibfnamefont {M.}~\bibnamefont {Hagan}}, \bibinfo {author} {\bibfnamefont
  {S.}~\bibnamefont {Bukkapatnam}}, \bibinfo {author} {\bibfnamefont
  {P.}~\bibnamefont {Agrawal}}, \ and\ \bibinfo {author} {\bibfnamefont
  {R.}~\bibnamefont {Komanduri}},\ }\bibfield  {title} {\enquote {\bibinfo
  {title} {Development of generalized potential-energy surfaces using many-body
  expansions, neural networks, and moiety energy approximations},}\ }\href@noop
  {} {\bibfield  {journal} {\bibinfo  {journal} {J. Chem. Phys.}\ }\textbf
  {\bibinfo {volume} {130}},\ \bibinfo {pages} {184102} (\bibinfo {year}
  {2009})}\BibitemShut {NoStop}%
\bibitem [{\citenamefont {Bart{\'o}k}\ \emph {et~al.}(2013)\citenamefont
  {Bart{\'o}k}, \citenamefont {Gillan}, \citenamefont {Manby},\ and\
  \citenamefont {Cs{\'a}nyi}}]{bartok2013machine}%
  \BibitemOpen
  \bibfield  {author} {\bibinfo {author} {\bibfnamefont {A.~P.}\ \bibnamefont
  {Bart{\'o}k}}, \bibinfo {author} {\bibfnamefont {M.~J.}\ \bibnamefont
  {Gillan}}, \bibinfo {author} {\bibfnamefont {F.~R.}\ \bibnamefont {Manby}}, \
  and\ \bibinfo {author} {\bibfnamefont {G.}~\bibnamefont {Cs{\'a}nyi}},\
  }\bibfield  {title} {\enquote {\bibinfo {title} {Machine-learning approach
  for one-and two-body corrections to density functional theory: Applications
  to molecular and condensed water},}\ }\href@noop {} {\bibfield  {journal}
  {\bibinfo  {journal} {Phys. Rev. B}\ }\textbf {\bibinfo {volume} {88}},\
  \bibinfo {pages} {054104} (\bibinfo {year} {2013})}\BibitemShut {NoStop}%
\bibitem [{\citenamefont {Cui}, \citenamefont {Liu},\ and\ \citenamefont
  {Jordan}(2006)}]{cui2006theoretical}%
  \BibitemOpen
  \bibfield  {author} {\bibinfo {author} {\bibfnamefont {J.}~\bibnamefont
  {Cui}}, \bibinfo {author} {\bibfnamefont {H.}~\bibnamefont {Liu}}, \ and\
  \bibinfo {author} {\bibfnamefont {K.~D.}\ \bibnamefont {Jordan}},\ }\bibfield
   {title} {\enquote {\bibinfo {title} {Theoretical characterization of the
  (h2o) 21 cluster: Application of an n-body decomposition procedure},}\
  }\href@noop {} {\bibfield  {journal} {\bibinfo  {journal} {J. Phys. Chem. B}\
  }\textbf {\bibinfo {volume} {110}},\ \bibinfo {pages} {18872--18878}
  (\bibinfo {year} {2006})}\BibitemShut {NoStop}%
\bibitem [{\citenamefont {G{\'o}ra}\ \emph {et~al.}(2011)\citenamefont
  {G{\'o}ra}, \citenamefont {Podeszwa}, \citenamefont {Cencek},\ and\
  \citenamefont {Szalewicz}}]{gora2011interaction}%
  \BibitemOpen
  \bibfield  {author} {\bibinfo {author} {\bibfnamefont {U.}~\bibnamefont
  {G{\'o}ra}}, \bibinfo {author} {\bibfnamefont {R.}~\bibnamefont {Podeszwa}},
  \bibinfo {author} {\bibfnamefont {W.}~\bibnamefont {Cencek}}, \ and\ \bibinfo
  {author} {\bibfnamefont {K.}~\bibnamefont {Szalewicz}},\ }\bibfield  {title}
  {\enquote {\bibinfo {title} {Interaction energies of large clusters from
  many-body expansion},}\ }\href@noop {} {\bibfield  {journal} {\bibinfo
  {journal} {J. Chem. Phys.}\ }\textbf {\bibinfo {volume} {135}},\ \bibinfo
  {pages} {224102} (\bibinfo {year} {2011})}\BibitemShut {NoStop}%
\bibitem [{\citenamefont {Hermann}\ \emph {et~al.}(2007)\citenamefont
  {Hermann}, \citenamefont {Krawczyk}, \citenamefont {Lein}, \citenamefont
  {Schwerdtfeger}, \citenamefont {Hamilton},\ and\ \citenamefont
  {Stewart}}]{hermann2007convergence}%
  \BibitemOpen
  \bibfield  {author} {\bibinfo {author} {\bibfnamefont {A.}~\bibnamefont
  {Hermann}}, \bibinfo {author} {\bibfnamefont {R.~P.}\ \bibnamefont
  {Krawczyk}}, \bibinfo {author} {\bibfnamefont {M.}~\bibnamefont {Lein}},
  \bibinfo {author} {\bibfnamefont {P.}~\bibnamefont {Schwerdtfeger}}, \bibinfo
  {author} {\bibfnamefont {I.~P.}\ \bibnamefont {Hamilton}}, \ and\ \bibinfo
  {author} {\bibfnamefont {J.~J.}\ \bibnamefont {Stewart}},\ }\bibfield
  {title} {\enquote {\bibinfo {title} {Convergence of the many-body expansion
  of interaction potentials: From van der waals to covalent and metallic
  systems},}\ }\href@noop {} {\bibfield  {journal} {\bibinfo  {journal} {Phys.
  Rev. A}\ }\textbf {\bibinfo {volume} {76}},\ \bibinfo {pages} {013202}
  (\bibinfo {year} {2007})}\BibitemShut {NoStop}%
\bibitem [{\citenamefont {Medders}\ and\ \citenamefont
  {Paesani}(2013)}]{medders2013many}%
  \BibitemOpen
  \bibfield  {author} {\bibinfo {author} {\bibfnamefont {G.~R.}\ \bibnamefont
  {Medders}}\ and\ \bibinfo {author} {\bibfnamefont {F.}~\bibnamefont
  {Paesani}},\ }\bibfield  {title} {\enquote {\bibinfo {title} {Many-body
  convergence of the electrostatic properties of water},}\ }\href@noop {}
  {\bibfield  {journal} {\bibinfo  {journal} {J. Chem. Theory Comput.}\
  }\textbf {\bibinfo {volume} {9}},\ \bibinfo {pages} {4844--4852} (\bibinfo
  {year} {2013})}\BibitemShut {NoStop}%
\bibitem [{\citenamefont {Paulus}\ \emph {et~al.}(2004)\citenamefont {Paulus},
  \citenamefont {Rosciszewski}, \citenamefont {Gaston}, \citenamefont
  {Schwerdtfeger},\ and\ \citenamefont {Stoll}}]{paulus2004convergence}%
  \BibitemOpen
  \bibfield  {author} {\bibinfo {author} {\bibfnamefont {B.}~\bibnamefont
  {Paulus}}, \bibinfo {author} {\bibfnamefont {K.}~\bibnamefont
  {Rosciszewski}}, \bibinfo {author} {\bibfnamefont {N.}~\bibnamefont
  {Gaston}}, \bibinfo {author} {\bibfnamefont {P.}~\bibnamefont
  {Schwerdtfeger}}, \ and\ \bibinfo {author} {\bibfnamefont {H.}~\bibnamefont
  {Stoll}},\ }\bibfield  {title} {\enquote {\bibinfo {title} {Convergence of
  the ab initio many-body expansion for the cohesive energy of solid
  mercury},}\ }\href@noop {} {\bibfield  {journal} {\bibinfo  {journal} {Phys.
  Rev. B}\ }\textbf {\bibinfo {volume} {70}},\ \bibinfo {pages} {165106}
  (\bibinfo {year} {2004})}\BibitemShut {NoStop}%
\bibitem [{\citenamefont {Richard}, \citenamefont {Lao},\ and\ \citenamefont
  {Herbert}(2014{\natexlab{a}})}]{richard2014aiming}%
  \BibitemOpen
  \bibfield  {author} {\bibinfo {author} {\bibfnamefont {R.~M.}\ \bibnamefont
  {Richard}}, \bibinfo {author} {\bibfnamefont {K.~U.}\ \bibnamefont {Lao}}, \
  and\ \bibinfo {author} {\bibfnamefont {J.~M.}\ \bibnamefont {Herbert}},\
  }\bibfield  {title} {\enquote {\bibinfo {title} {Aiming for benchmark
  accuracy with the many-body expansion},}\ }\href@noop {} {\bibfield
  {journal} {\bibinfo  {journal} {Accounts Chem. Res.}\ }\textbf {\bibinfo
  {volume} {47}},\ \bibinfo {pages} {2828--2836} (\bibinfo {year}
  {2014}{\natexlab{a}})}\BibitemShut {NoStop}%
\bibitem [{\citenamefont {Richard}, \citenamefont {Lao},\ and\ \citenamefont
  {Herbert}(2014{\natexlab{b}})}]{richard2014understanding}%
  \BibitemOpen
  \bibfield  {author} {\bibinfo {author} {\bibfnamefont {R.~M.}\ \bibnamefont
  {Richard}}, \bibinfo {author} {\bibfnamefont {K.~U.}\ \bibnamefont {Lao}}, \
  and\ \bibinfo {author} {\bibfnamefont {J.~M.}\ \bibnamefont {Herbert}},\
  }\bibfield  {title} {\enquote {\bibinfo {title} {Understanding the many-body
  expansion for large systems. i. precision considerations},}\ }\href@noop {}
  {\bibfield  {journal} {\bibinfo  {journal} {J. Chem. Phys.}\ }\textbf
  {\bibinfo {volume} {141}},\ \bibinfo {pages} {014108} (\bibinfo {year}
  {2014}{\natexlab{b}})}\BibitemShut {NoStop}%
\bibitem [{\citenamefont {Fennell}\ and\ \citenamefont
  {Gezelter}(2006)}]{fennell2006ewald}%
  \BibitemOpen
  \bibfield  {author} {\bibinfo {author} {\bibfnamefont {C.~J.}\ \bibnamefont
  {Fennell}}\ and\ \bibinfo {author} {\bibfnamefont {J.~D.}\ \bibnamefont
  {Gezelter}},\ }\bibfield  {title} {\enquote {\bibinfo {title} {Is the ewald
  summation still necessary? pairwise alternatives to the accepted standard for
  long-range electrostatics},}\ }\href@noop {} {\bibfield  {journal} {\bibinfo
  {journal} {J. Chem. Phys.}\ }\textbf {\bibinfo {volume} {124}},\ \bibinfo
  {pages} {234104} (\bibinfo {year} {2006})}\BibitemShut {NoStop}%
\bibitem [{\citenamefont {Lamichhane}, \citenamefont {Gezelter},\ and\
  \citenamefont {Newman}(2014)}]{lamichhane2014real1}%
  \BibitemOpen
  \bibfield  {author} {\bibinfo {author} {\bibfnamefont {M.}~\bibnamefont
  {Lamichhane}}, \bibinfo {author} {\bibfnamefont {J.~D.}\ \bibnamefont
  {Gezelter}}, \ and\ \bibinfo {author} {\bibfnamefont {K.~E.}\ \bibnamefont
  {Newman}},\ }\bibfield  {title} {\enquote {\bibinfo {title} {Real space
  electrostatics for multipoles. i. development of methods},}\ }\href@noop {}
  {\bibfield  {journal} {\bibinfo  {journal} {J. Chem. Phys.}\ }\textbf
  {\bibinfo {volume} {141}},\ \bibinfo {pages} {134109} (\bibinfo {year}
  {2014})}\BibitemShut {NoStop}%
\bibitem [{\citenamefont {Lamichhane}, \citenamefont {Newman},\ and\
  \citenamefont {Gezelter}(2014)}]{lamichhane2014real2}%
  \BibitemOpen
  \bibfield  {author} {\bibinfo {author} {\bibfnamefont {M.}~\bibnamefont
  {Lamichhane}}, \bibinfo {author} {\bibfnamefont {K.~E.}\ \bibnamefont
  {Newman}}, \ and\ \bibinfo {author} {\bibfnamefont {J.~D.}\ \bibnamefont
  {Gezelter}},\ }\bibfield  {title} {\enquote {\bibinfo {title} {Real space
  electrostatics for multipoles. ii. comparisons with the ewald sum},}\
  }\href@noop {} {\bibfield  {journal} {\bibinfo  {journal} {J. Chem. Phys.}\
  }\textbf {\bibinfo {volume} {141}},\ \bibinfo {pages} {134110} (\bibinfo
  {year} {2014})}\BibitemShut {NoStop}%
\bibitem [{\citenamefont {D.A.~Case}\ and\ \citenamefont {Kollman}()}]{Amber}%
  \BibitemOpen
  \bibfield  {author} {\bibinfo {author} {\bibfnamefont {D.~C. T. C. I. T. D.
  R. D. T. G. H. G. A. G. N. H. S. I. P. J. J. K. A. K. T. L. S. L. P. L. C. L.
  T. L. R. L. B. M. D. M. K. M. G. M. H. N. H. N. I. O. A. O. D. R. A. R. C. S.
  C. S. W. B.-S. J. S. R. W. J. W. R. W. X. W. L.~X.}\ \bibnamefont
  {D.A.~Case}, \bibfnamefont {R.M.~Betz}}\ and\ \bibinfo {author}
  {\bibfnamefont {P.}~\bibnamefont {Kollman}},\ }\href@noop {} {\enquote
  {\bibinfo {title} {Amber 2016, university of california, san francisco.}}\
  }\BibitemShut {NoStop}%
\bibitem [{\citenamefont {Shao}\ \emph {et~al.}(2015)\citenamefont {Shao},
  \citenamefont {Gan}, \citenamefont {Epifanovsky}, \citenamefont {Gilbert},
  \citenamefont {Wormit}, \citenamefont {Kussmann}, \citenamefont {Lange},
  \citenamefont {Behn}, \citenamefont {Deng}, \citenamefont {Feng},
  \citenamefont {Ghosh}, \citenamefont {Goldey}, \citenamefont {Horn},
  \citenamefont {Jacobson}, \citenamefont {Kaliman}, \citenamefont
  {Khaliullin}, \citenamefont {Ku{\'s}}, \citenamefont {Landau}, \citenamefont
  {Liu}, \citenamefont {Proynov}, \citenamefont {Rhee}, \citenamefont
  {Richard}, \citenamefont {Rohrdanz}, \citenamefont {Steele}, \citenamefont
  {Sundstrom}, \citenamefont {Woodcock}, \citenamefont {Zimmerman},
  \citenamefont {Zuev}, \citenamefont {Albrecht}, \citenamefont {Alguire},
  \citenamefont {Austin}, \citenamefont {Beran}, \citenamefont {Bernard},
  \citenamefont {Berquist}, \citenamefont {Brandhorst}, \citenamefont
  {Bravaya}, \citenamefont {Brown}, \citenamefont {Casanova}, \citenamefont
  {Chang}, \citenamefont {Chen}, \citenamefont {Chien}, \citenamefont
  {Closser}, \citenamefont {Crittenden}, \citenamefont {Diedenhofen},
  \citenamefont {DiStasio}, \citenamefont {Do}, \citenamefont {Dutoi},
  \citenamefont {Edgar}, \citenamefont {Fatehi}, \citenamefont {Fusti-Molnar},
  \citenamefont {Ghysels}, \citenamefont {Golubeva-Zadorozhnaya}, \citenamefont
  {Gomes}, \citenamefont {Hanson-Heine}, \citenamefont {Harbach}, \citenamefont
  {Hauser}, \citenamefont {Hohenstein}, \citenamefont {Holden}, \citenamefont
  {Jagau}, \citenamefont {Ji}, \citenamefont {Kaduk}, \citenamefont
  {Khistyaev}, \citenamefont {Kim}, \citenamefont {Kim}, \citenamefont {King},
  \citenamefont {Klunzinger}, \citenamefont {Kosenkov}, \citenamefont
  {Kowalczyk}, \citenamefont {Krauter}, \citenamefont {Lao}, \citenamefont
  {Laurent}, \citenamefont {Lawler}, \citenamefont {Levchenko}, \citenamefont
  {Lin}, \citenamefont {Liu}, \citenamefont {Livshits}, \citenamefont {Lochan},
  \citenamefont {Luenser}, \citenamefont {Manohar}, \citenamefont {Manzer},
  \citenamefont {Mao}, \citenamefont {Mardirossian}, \citenamefont {Marenich},
  \citenamefont {Maurer}, \citenamefont {Mayhall}, \citenamefont {Neuscamman},
  \citenamefont {Oana}, \citenamefont {Olivares-Amaya}, \citenamefont
  {O'Neill}, \citenamefont {Parkhill}, \citenamefont {Perrine}, \citenamefont
  {Peverati}, \citenamefont {Prociuk}, \citenamefont {Rehn}, \citenamefont
  {Rosta}, \citenamefont {Russ}, \citenamefont {Sharada}, \citenamefont
  {Sharma}, \citenamefont {Small}, \citenamefont {Sodt}, \citenamefont {Stein},
  \citenamefont {St{\"u}ck}, \citenamefont {Su}, \citenamefont {Thom},
  \citenamefont {Tsuchimochi}, \citenamefont {Vanovschi}, \citenamefont {Vogt},
  \citenamefont {Vydrov}, \citenamefont {Wang}, \citenamefont {Watson},
  \citenamefont {Wenzel}, \citenamefont {White}, \citenamefont {Williams},
  \citenamefont {Yang}, \citenamefont {Yeganeh}, \citenamefont {Yost},
  \citenamefont {You}, \citenamefont {Zhang}, \citenamefont {Zhang},
  \citenamefont {Zhao}, \citenamefont {Brooks}, \citenamefont {Chan},
  \citenamefont {Chipman}, \citenamefont {Cramer}, \citenamefont {Goddard},
  \citenamefont {Gordon}, \citenamefont {Hehre}, \citenamefont {Klamt},
  \citenamefont {Schaefer}, \citenamefont {Schmidt}, \citenamefont {Sherrill},
  \citenamefont {Truhlar}, \citenamefont {Warshel}, \citenamefont {Xu},
  \citenamefont {Aspuru-Guzik}, \citenamefont {Baer}, \citenamefont {Bell},
  \citenamefont {Besley}, \citenamefont {Chai}, \citenamefont {Dreuw},
  \citenamefont {Dunietz}, \citenamefont {Furlani}, \citenamefont {Gwaltney},
  \citenamefont {Hsu}, \citenamefont {Jung}, \citenamefont {Kong},
  \citenamefont {Lambrecht}, \citenamefont {Liang}, \citenamefont {Ochsenfeld},
  \citenamefont {Rassolov}, \citenamefont {Slipchenko}, \citenamefont
  {Subotnik}, \citenamefont {Van~Voorhis}, \citenamefont {Herbert},
  \citenamefont {Krylov}, \citenamefont {Gill},\ and\ \citenamefont
  {Head-Gordon}}]{shao2015advances}%
  \BibitemOpen
  \bibfield  {author} {\bibinfo {author} {\bibfnamefont {Y.}~\bibnamefont
  {Shao}}, \bibinfo {author} {\bibfnamefont {Z.}~\bibnamefont {Gan}}, \bibinfo
  {author} {\bibfnamefont {E.}~\bibnamefont {Epifanovsky}}, \bibinfo {author}
  {\bibfnamefont {A.~T.}\ \bibnamefont {Gilbert}}, \bibinfo {author}
  {\bibfnamefont {M.}~\bibnamefont {Wormit}}, \bibinfo {author} {\bibfnamefont
  {J.}~\bibnamefont {Kussmann}}, \bibinfo {author} {\bibfnamefont {A.~W.}\
  \bibnamefont {Lange}}, \bibinfo {author} {\bibfnamefont {A.}~\bibnamefont
  {Behn}}, \bibinfo {author} {\bibfnamefont {J.}~\bibnamefont {Deng}}, \bibinfo
  {author} {\bibfnamefont {X.}~\bibnamefont {Feng}}, \bibinfo {author}
  {\bibfnamefont {D.}~\bibnamefont {Ghosh}}, \bibinfo {author} {\bibfnamefont
  {M.}~\bibnamefont {Goldey}}, \bibinfo {author} {\bibfnamefont {P.~R.}\
  \bibnamefont {Horn}}, \bibinfo {author} {\bibfnamefont {L.~D.}\ \bibnamefont
  {Jacobson}}, \bibinfo {author} {\bibfnamefont {I.}~\bibnamefont {Kaliman}},
  \bibinfo {author} {\bibfnamefont {R.~Z.}\ \bibnamefont {Khaliullin}},
  \bibinfo {author} {\bibfnamefont {T.}~\bibnamefont {Ku{\'s}}}, \bibinfo
  {author} {\bibfnamefont {A.}~\bibnamefont {Landau}}, \bibinfo {author}
  {\bibfnamefont {J.}~\bibnamefont {Liu}}, \bibinfo {author} {\bibfnamefont
  {E.~I.}\ \bibnamefont {Proynov}}, \bibinfo {author} {\bibfnamefont {Y.~M.}\
  \bibnamefont {Rhee}}, \bibinfo {author} {\bibfnamefont {R.~M.}\ \bibnamefont
  {Richard}}, \bibinfo {author} {\bibfnamefont {M.~A.}\ \bibnamefont
  {Rohrdanz}}, \bibinfo {author} {\bibfnamefont {R.~P.}\ \bibnamefont
  {Steele}}, \bibinfo {author} {\bibfnamefont {E.~J.}\ \bibnamefont
  {Sundstrom}}, \bibinfo {author} {\bibfnamefont {H.~L.}\ \bibnamefont
  {Woodcock}}, \bibinfo {author} {\bibfnamefont {P.~M.}\ \bibnamefont
  {Zimmerman}}, \bibinfo {author} {\bibfnamefont {D.}~\bibnamefont {Zuev}},
  \bibinfo {author} {\bibfnamefont {B.}~\bibnamefont {Albrecht}}, \bibinfo
  {author} {\bibfnamefont {E.}~\bibnamefont {Alguire}}, \bibinfo {author}
  {\bibfnamefont {B.}~\bibnamefont {Austin}}, \bibinfo {author} {\bibfnamefont
  {G.~J.~O.}\ \bibnamefont {Beran}}, \bibinfo {author} {\bibfnamefont {Y.~A.}\
  \bibnamefont {Bernard}}, \bibinfo {author} {\bibfnamefont {E.}~\bibnamefont
  {Berquist}}, \bibinfo {author} {\bibfnamefont {K.}~\bibnamefont
  {Brandhorst}}, \bibinfo {author} {\bibfnamefont {K.~B.}\ \bibnamefont
  {Bravaya}}, \bibinfo {author} {\bibfnamefont {S.~T.}\ \bibnamefont {Brown}},
  \bibinfo {author} {\bibfnamefont {D.}~\bibnamefont {Casanova}}, \bibinfo
  {author} {\bibfnamefont {C.-M.}\ \bibnamefont {Chang}}, \bibinfo {author}
  {\bibfnamefont {Y.}~\bibnamefont {Chen}}, \bibinfo {author} {\bibfnamefont
  {S.~H.}\ \bibnamefont {Chien}}, \bibinfo {author} {\bibfnamefont {K.~D.}\
  \bibnamefont {Closser}}, \bibinfo {author} {\bibfnamefont {D.~L.}\
  \bibnamefont {Crittenden}}, \bibinfo {author} {\bibfnamefont
  {M.}~\bibnamefont {Diedenhofen}}, \bibinfo {author} {\bibfnamefont {R.~A.}\
  \bibnamefont {DiStasio}}, \bibinfo {author} {\bibfnamefont {H.}~\bibnamefont
  {Do}}, \bibinfo {author} {\bibfnamefont {A.~D.}\ \bibnamefont {Dutoi}},
  \bibinfo {author} {\bibfnamefont {R.~G.}\ \bibnamefont {Edgar}}, \bibinfo
  {author} {\bibfnamefont {S.}~\bibnamefont {Fatehi}}, \bibinfo {author}
  {\bibfnamefont {L.}~\bibnamefont {Fusti-Molnar}}, \bibinfo {author}
  {\bibfnamefont {A.}~\bibnamefont {Ghysels}}, \bibinfo {author} {\bibfnamefont
  {A.}~\bibnamefont {Golubeva-Zadorozhnaya}}, \bibinfo {author} {\bibfnamefont
  {J.}~\bibnamefont {Gomes}}, \bibinfo {author} {\bibfnamefont {M.~W.}\
  \bibnamefont {Hanson-Heine}}, \bibinfo {author} {\bibfnamefont {P.~H.}\
  \bibnamefont {Harbach}}, \bibinfo {author} {\bibfnamefont {A.~W.}\
  \bibnamefont {Hauser}}, \bibinfo {author} {\bibfnamefont {E.~G.}\
  \bibnamefont {Hohenstein}}, \bibinfo {author} {\bibfnamefont {Z.~C.}\
  \bibnamefont {Holden}}, \bibinfo {author} {\bibfnamefont {T.-C.}\
  \bibnamefont {Jagau}}, \bibinfo {author} {\bibfnamefont {H.}~\bibnamefont
  {Ji}}, \bibinfo {author} {\bibfnamefont {B.}~\bibnamefont {Kaduk}}, \bibinfo
  {author} {\bibfnamefont {K.}~\bibnamefont {Khistyaev}}, \bibinfo {author}
  {\bibfnamefont {J.}~\bibnamefont {Kim}}, \bibinfo {author} {\bibfnamefont
  {J.}~\bibnamefont {Kim}}, \bibinfo {author} {\bibfnamefont {R.~A.}\
  \bibnamefont {King}}, \bibinfo {author} {\bibfnamefont {P.}~\bibnamefont
  {Klunzinger}}, \bibinfo {author} {\bibfnamefont {D.}~\bibnamefont
  {Kosenkov}}, \bibinfo {author} {\bibfnamefont {T.}~\bibnamefont {Kowalczyk}},
  \bibinfo {author} {\bibfnamefont {C.~M.}\ \bibnamefont {Krauter}}, \bibinfo
  {author} {\bibfnamefont {K.~U.}\ \bibnamefont {Lao}}, \bibinfo {author}
  {\bibfnamefont {A.}~\bibnamefont {Laurent}}, \bibinfo {author} {\bibfnamefont
  {K.~V.}\ \bibnamefont {Lawler}}, \bibinfo {author} {\bibfnamefont {S.~V.}\
  \bibnamefont {Levchenko}}, \bibinfo {author} {\bibfnamefont {C.~Y.}\
  \bibnamefont {Lin}}, \bibinfo {author} {\bibfnamefont {F.}~\bibnamefont
  {Liu}}, \bibinfo {author} {\bibfnamefont {E.}~\bibnamefont {Livshits}},
  \bibinfo {author} {\bibfnamefont {R.~C.}\ \bibnamefont {Lochan}}, \bibinfo
  {author} {\bibfnamefont {A.}~\bibnamefont {Luenser}}, \bibinfo {author}
  {\bibfnamefont {P.}~\bibnamefont {Manohar}}, \bibinfo {author} {\bibfnamefont
  {S.~F.}\ \bibnamefont {Manzer}}, \bibinfo {author} {\bibfnamefont {S.-P.}\
  \bibnamefont {Mao}}, \bibinfo {author} {\bibfnamefont {N.}~\bibnamefont
  {Mardirossian}}, \bibinfo {author} {\bibfnamefont {A.~V.}\ \bibnamefont
  {Marenich}}, \bibinfo {author} {\bibfnamefont {S.~A.}\ \bibnamefont
  {Maurer}}, \bibinfo {author} {\bibfnamefont {N.~J.}\ \bibnamefont {Mayhall}},
  \bibinfo {author} {\bibfnamefont {E.}~\bibnamefont {Neuscamman}}, \bibinfo
  {author} {\bibfnamefont {C.~M.}\ \bibnamefont {Oana}}, \bibinfo {author}
  {\bibfnamefont {R.}~\bibnamefont {Olivares-Amaya}}, \bibinfo {author}
  {\bibfnamefont {D.~P.}\ \bibnamefont {O'Neill}}, \bibinfo {author}
  {\bibfnamefont {J.~A.}\ \bibnamefont {Parkhill}}, \bibinfo {author}
  {\bibfnamefont {T.~M.}\ \bibnamefont {Perrine}}, \bibinfo {author}
  {\bibfnamefont {R.}~\bibnamefont {Peverati}}, \bibinfo {author}
  {\bibfnamefont {A.}~\bibnamefont {Prociuk}}, \bibinfo {author} {\bibfnamefont
  {D.~R.}\ \bibnamefont {Rehn}}, \bibinfo {author} {\bibfnamefont
  {E.}~\bibnamefont {Rosta}}, \bibinfo {author} {\bibfnamefont {N.~J.}\
  \bibnamefont {Russ}}, \bibinfo {author} {\bibfnamefont {S.~M.}\ \bibnamefont
  {Sharada}}, \bibinfo {author} {\bibfnamefont {S.}~\bibnamefont {Sharma}},
  \bibinfo {author} {\bibfnamefont {D.~W.}\ \bibnamefont {Small}}, \bibinfo
  {author} {\bibfnamefont {A.}~\bibnamefont {Sodt}}, \bibinfo {author}
  {\bibfnamefont {T.}~\bibnamefont {Stein}}, \bibinfo {author} {\bibfnamefont
  {D.}~\bibnamefont {St{\"u}ck}}, \bibinfo {author} {\bibfnamefont {Y.-C.}\
  \bibnamefont {Su}}, \bibinfo {author} {\bibfnamefont {A.~J.}\ \bibnamefont
  {Thom}}, \bibinfo {author} {\bibfnamefont {T.}~\bibnamefont {Tsuchimochi}},
  \bibinfo {author} {\bibfnamefont {V.}~\bibnamefont {Vanovschi}}, \bibinfo
  {author} {\bibfnamefont {L.}~\bibnamefont {Vogt}}, \bibinfo {author}
  {\bibfnamefont {O.}~\bibnamefont {Vydrov}}, \bibinfo {author} {\bibfnamefont
  {T.}~\bibnamefont {Wang}}, \bibinfo {author} {\bibfnamefont {M.~A.}\
  \bibnamefont {Watson}}, \bibinfo {author} {\bibfnamefont {J.}~\bibnamefont
  {Wenzel}}, \bibinfo {author} {\bibfnamefont {A.}~\bibnamefont {White}},
  \bibinfo {author} {\bibfnamefont {C.~F.}\ \bibnamefont {Williams}}, \bibinfo
  {author} {\bibfnamefont {J.}~\bibnamefont {Yang}}, \bibinfo {author}
  {\bibfnamefont {S.}~\bibnamefont {Yeganeh}}, \bibinfo {author} {\bibfnamefont
  {S.~R.}\ \bibnamefont {Yost}}, \bibinfo {author} {\bibfnamefont {Z.-Q.}\
  \bibnamefont {You}}, \bibinfo {author} {\bibfnamefont {I.~Y.}\ \bibnamefont
  {Zhang}}, \bibinfo {author} {\bibfnamefont {X.}~\bibnamefont {Zhang}},
  \bibinfo {author} {\bibfnamefont {Y.}~\bibnamefont {Zhao}}, \bibinfo {author}
  {\bibfnamefont {B.~R.}\ \bibnamefont {Brooks}}, \bibinfo {author}
  {\bibfnamefont {G.~K.}\ \bibnamefont {Chan}}, \bibinfo {author}
  {\bibfnamefont {D.~M.}\ \bibnamefont {Chipman}}, \bibinfo {author}
  {\bibfnamefont {C.~J.}\ \bibnamefont {Cramer}}, \bibinfo {author}
  {\bibfnamefont {W.~A.}\ \bibnamefont {Goddard}}, \bibinfo {author}
  {\bibfnamefont {M.~S.}\ \bibnamefont {Gordon}}, \bibinfo {author}
  {\bibfnamefont {W.~J.}\ \bibnamefont {Hehre}}, \bibinfo {author}
  {\bibfnamefont {A.}~\bibnamefont {Klamt}}, \bibinfo {author} {\bibfnamefont
  {H.~F.}\ \bibnamefont {Schaefer}}, \bibinfo {author} {\bibfnamefont {M.~W.}\
  \bibnamefont {Schmidt}}, \bibinfo {author} {\bibfnamefont {C.~D.}\
  \bibnamefont {Sherrill}}, \bibinfo {author} {\bibfnamefont {D.~G.}\
  \bibnamefont {Truhlar}}, \bibinfo {author} {\bibfnamefont {A.}~\bibnamefont
  {Warshel}}, \bibinfo {author} {\bibfnamefont {X.}~\bibnamefont {Xu}},
  \bibinfo {author} {\bibfnamefont {A.}~\bibnamefont {Aspuru-Guzik}}, \bibinfo
  {author} {\bibfnamefont {R.}~\bibnamefont {Baer}}, \bibinfo {author}
  {\bibfnamefont {A.~T.}\ \bibnamefont {Bell}}, \bibinfo {author}
  {\bibfnamefont {N.~A.}\ \bibnamefont {Besley}}, \bibinfo {author}
  {\bibfnamefont {J.-D.}\ \bibnamefont {Chai}}, \bibinfo {author}
  {\bibfnamefont {A.}~\bibnamefont {Dreuw}}, \bibinfo {author} {\bibfnamefont
  {B.~D.}\ \bibnamefont {Dunietz}}, \bibinfo {author} {\bibfnamefont {T.~R.}\
  \bibnamefont {Furlani}}, \bibinfo {author} {\bibfnamefont {S.~R.}\
  \bibnamefont {Gwaltney}}, \bibinfo {author} {\bibfnamefont {C.-P.}\
  \bibnamefont {Hsu}}, \bibinfo {author} {\bibfnamefont {Y.}~\bibnamefont
  {Jung}}, \bibinfo {author} {\bibfnamefont {J.}~\bibnamefont {Kong}}, \bibinfo
  {author} {\bibfnamefont {D.~S.}\ \bibnamefont {Lambrecht}}, \bibinfo {author}
  {\bibfnamefont {W.}~\bibnamefont {Liang}}, \bibinfo {author} {\bibfnamefont
  {C.}~\bibnamefont {Ochsenfeld}}, \bibinfo {author} {\bibfnamefont {V.~A.}\
  \bibnamefont {Rassolov}}, \bibinfo {author} {\bibfnamefont {L.~V.}\
  \bibnamefont {Slipchenko}}, \bibinfo {author} {\bibfnamefont {J.~E.}\
  \bibnamefont {Subotnik}}, \bibinfo {author} {\bibfnamefont {T.}~\bibnamefont
  {Van~Voorhis}}, \bibinfo {author} {\bibfnamefont {J.~M.}\ \bibnamefont
  {Herbert}}, \bibinfo {author} {\bibfnamefont {A.~I.}\ \bibnamefont {Krylov}},
  \bibinfo {author} {\bibfnamefont {P.~M.}\ \bibnamefont {Gill}}, \ and\
  \bibinfo {author} {\bibfnamefont {M.}~\bibnamefont {Head-Gordon}},\
  }\bibfield  {title} {\enquote {\bibinfo {title} {Advances in molecular
  quantum chemistry contained in the q-chem 4 program package},}\ }\href
  {\doibase 10.1080/00268976.2014.952696} {\bibfield  {journal} {\bibinfo
  {journal} {Mol. Phys.}\ }\textbf {\bibinfo {volume} {113}},\ \bibinfo {pages}
  {184--215} (\bibinfo {year} {2015})}\BibitemShut {NoStop}%
\bibitem [{\citenamefont {Krizhevsky}, \citenamefont {Sutskever},\ and\
  \citenamefont {Hinton}(2012)}]{krizhevsky2012imagenet}%
  \BibitemOpen
  \bibfield  {author} {\bibinfo {author} {\bibfnamefont {A.}~\bibnamefont
  {Krizhevsky}}, \bibinfo {author} {\bibfnamefont {I.}~\bibnamefont
  {Sutskever}}, \ and\ \bibinfo {author} {\bibfnamefont {G.~E.}\ \bibnamefont
  {Hinton}},\ }\bibfield  {title} {\enquote {\bibinfo {title} {Imagenet
  classification with deep convolutional neural networks},}\ }in\ \href
  {http://papers.nips.cc/paper/4824-imagenet-classification-with-deep-convolutional-neural-networks.pdf}
  {\emph {\bibinfo {booktitle} {Adv. Neural Inf. Process. Syst. 25}}},\
  \bibinfo {editor} {edited by\ \bibinfo {editor} {\bibfnamefont
  {F.}~\bibnamefont {Pereira}}, \bibinfo {editor} {\bibfnamefont {C.~J.~C.}\
  \bibnamefont {Burges}}, \bibinfo {editor} {\bibfnamefont {L.}~\bibnamefont
  {Bottou}}, \ and\ \bibinfo {editor} {\bibfnamefont {K.~Q.}\ \bibnamefont
  {Weinberger}}}\ (\bibinfo  {publisher} {Curran Associates, Inc.},\ \bibinfo
  {year} {2012})\ pp.\ \bibinfo {pages} {1097--1105}\BibitemShut {NoStop}%
\bibitem [{\citenamefont {Hansen}\ \emph {et~al.}(2013)\citenamefont {Hansen},
  \citenamefont {Montavon}, \citenamefont {Biegler}, \citenamefont {Fazli},
  \citenamefont {Rupp}, \citenamefont {Scheffler}, \citenamefont
  {Von~Lilienfeld}, \citenamefont {Tkatchenko},\ and\ \citenamefont
  {Müller}}]{hansen2013assessment}%
  \BibitemOpen
  \bibfield  {author} {\bibinfo {author} {\bibfnamefont {K.}~\bibnamefont
  {Hansen}}, \bibinfo {author} {\bibfnamefont {G.}~\bibnamefont {Montavon}},
  \bibinfo {author} {\bibfnamefont {F.}~\bibnamefont {Biegler}}, \bibinfo
  {author} {\bibfnamefont {S.}~\bibnamefont {Fazli}}, \bibinfo {author}
  {\bibfnamefont {M.}~\bibnamefont {Rupp}}, \bibinfo {author} {\bibfnamefont
  {M.}~\bibnamefont {Scheffler}}, \bibinfo {author} {\bibfnamefont {O.~A.}\
  \bibnamefont {Von~Lilienfeld}}, \bibinfo {author} {\bibfnamefont
  {A.}~\bibnamefont {Tkatchenko}}, \ and\ \bibinfo {author} {\bibfnamefont
  {K.-R.}\ \bibnamefont {Müller}},\ }\bibfield  {title} {\enquote {\bibinfo
  {title} {Assessment and validation of machine learning methods for predicting
  molecular atomization energies},}\ }\href@noop {} {\bibfield  {journal}
  {\bibinfo  {journal} {J. Chem. Theory Comput.}\ }\textbf {\bibinfo {volume}
  {9}},\ \bibinfo {pages} {3404--3419} (\bibinfo {year} {2013})}\BibitemShut
  {NoStop}%
\bibitem [{\citenamefont {Montavon}\ \emph {et~al.}(2012)\citenamefont
  {Montavon}, \citenamefont {Hansen}, \citenamefont {Fazli}, \citenamefont
  {Rupp}, \citenamefont {Biegler}, \citenamefont {Ziehe}, \citenamefont
  {Tkatchenko}, \citenamefont {Lilienfeld},\ and\ \citenamefont
  {M\"{u}ller}}]{montavon2012learning}%
  \BibitemOpen
  \bibfield  {author} {\bibinfo {author} {\bibfnamefont {G.}~\bibnamefont
  {Montavon}}, \bibinfo {author} {\bibfnamefont {K.}~\bibnamefont {Hansen}},
  \bibinfo {author} {\bibfnamefont {S.}~\bibnamefont {Fazli}}, \bibinfo
  {author} {\bibfnamefont {M.}~\bibnamefont {Rupp}}, \bibinfo {author}
  {\bibfnamefont {F.}~\bibnamefont {Biegler}}, \bibinfo {author} {\bibfnamefont
  {A.}~\bibnamefont {Ziehe}}, \bibinfo {author} {\bibfnamefont
  {A.}~\bibnamefont {Tkatchenko}}, \bibinfo {author} {\bibfnamefont {A.~V.}\
  \bibnamefont {Lilienfeld}}, \ and\ \bibinfo {author} {\bibfnamefont {K.-R.}\
  \bibnamefont {M\"{u}ller}},\ }\bibfield  {title} {\enquote {\bibinfo {title}
  {Learning invariant representations of molecules for atomization energy
  prediction},}\ }in\ \href
  {http://papers.nips.cc/paper/4830-learning-invariant-representations-of-molecules-for-atomization-energy-prediction.pdf}
  {\emph {\bibinfo {booktitle} {Adv. Neural Inf. Process. Syst. 25}}},\
  \bibinfo {editor} {edited by\ \bibinfo {editor} {\bibfnamefont
  {F.}~\bibnamefont {Pereira}}, \bibinfo {editor} {\bibfnamefont {C.~J.~C.}\
  \bibnamefont {Burges}}, \bibinfo {editor} {\bibfnamefont {L.}~\bibnamefont
  {Bottou}}, \ and\ \bibinfo {editor} {\bibfnamefont {K.~Q.}\ \bibnamefont
  {Weinberger}}}\ (\bibinfo  {publisher} {Curran Associates, Inc.},\ \bibinfo
  {year} {2012})\ pp.\ \bibinfo {pages} {440--448}\BibitemShut {NoStop}%
\bibitem [{\citenamefont {Faber}\ \emph {et~al.}(2015)\citenamefont {Faber},
  \citenamefont {Lindmaa}, \citenamefont {von Lilienfeld},\ and\ \citenamefont
  {Armiento}}]{faber2015crystal}%
  \BibitemOpen
  \bibfield  {author} {\bibinfo {author} {\bibfnamefont {F.}~\bibnamefont
  {Faber}}, \bibinfo {author} {\bibfnamefont {A.}~\bibnamefont {Lindmaa}},
  \bibinfo {author} {\bibfnamefont {O.~A.}\ \bibnamefont {von Lilienfeld}}, \
  and\ \bibinfo {author} {\bibfnamefont {R.}~\bibnamefont {Armiento}},\
  }\bibfield  {title} {\enquote {\bibinfo {title} {Crystal structure
  representations for machine learning models of formation energies},}\
  }\href@noop {} {\bibfield  {journal} {\bibinfo  {journal} {Int. J. Quantum
  Chem.}\ }\textbf {\bibinfo {volume} {115}},\ \bibinfo {pages} {1094--1101}
  (\bibinfo {year} {2015})}\BibitemShut {NoStop}%
\bibitem [{\citenamefont {Bart{\'o}k}\ and\ \citenamefont
  {Payne}(2010)}]{bartok2010gaussian}%
  \BibitemOpen
  \bibfield  {author} {\bibinfo {author} {\bibfnamefont {A.~P.}\ \bibnamefont
  {Bart{\'o}k}}\ and\ \bibinfo {author} {\bibfnamefont {M.~C.}\ \bibnamefont
  {Payne}},\ }\bibfield  {title} {\enquote {\bibinfo {title} {Gaussian
  approximation potentials: The accuracy of quantum mechanics, without the
  electrons},}\ }\href@noop {} {\bibfield  {journal} {\bibinfo  {journal}
  {Phys. Rev. Lett.}\ }\textbf {\bibinfo {volume} {104}},\ \bibinfo {pages}
  {136403} (\bibinfo {year} {2010})}\BibitemShut {NoStop}%
\bibitem [{\citenamefont {Bart{\'o}k}, \citenamefont {Kondor},\ and\
  \citenamefont {Cs{\'a}nyi}(2013)}]{bartok2013representing}%
  \BibitemOpen
  \bibfield  {author} {\bibinfo {author} {\bibfnamefont {A.~P.}\ \bibnamefont
  {Bart{\'o}k}}, \bibinfo {author} {\bibfnamefont {R.}~\bibnamefont {Kondor}},
  \ and\ \bibinfo {author} {\bibfnamefont {G.}~\bibnamefont {Cs{\'a}nyi}},\
  }\bibfield  {title} {\enquote {\bibinfo {title} {On representing chemical
  environments},}\ }\href@noop {} {\bibfield  {journal} {\bibinfo  {journal}
  {Phys. Rev. B}\ }\textbf {\bibinfo {volume} {87}},\ \bibinfo {pages} {184115}
  (\bibinfo {year} {2013})}\BibitemShut {NoStop}%
\bibitem [{\citenamefont {Sadeghi}\ \emph {et~al.}(2013)\citenamefont
  {Sadeghi}, \citenamefont {Ghasemi}, \citenamefont {Schaefer}, \citenamefont
  {Mohr}, \citenamefont {Lill},\ and\ \citenamefont
  {Goedecker}}]{sadeghi2013metrics}%
  \BibitemOpen
  \bibfield  {author} {\bibinfo {author} {\bibfnamefont {A.}~\bibnamefont
  {Sadeghi}}, \bibinfo {author} {\bibfnamefont {S.~A.}\ \bibnamefont
  {Ghasemi}}, \bibinfo {author} {\bibfnamefont {B.}~\bibnamefont {Schaefer}},
  \bibinfo {author} {\bibfnamefont {S.}~\bibnamefont {Mohr}}, \bibinfo {author}
  {\bibfnamefont {M.~A.}\ \bibnamefont {Lill}}, \ and\ \bibinfo {author}
  {\bibfnamefont {S.}~\bibnamefont {Goedecker}},\ }\bibfield  {title} {\enquote
  {\bibinfo {title} {Metrics for measuring distances in configuration
  spaces},}\ }\href@noop {} {\bibfield  {journal} {\bibinfo  {journal} {J.
  Chem. Phys.}\ }\textbf {\bibinfo {volume} {139}},\ \bibinfo {pages} {184118}
  (\bibinfo {year} {2013})}\BibitemShut {NoStop}%
\bibitem [{\citenamefont {Zhu}\ \emph {et~al.}(2016)\citenamefont {Zhu},
  \citenamefont {Amsler}, \citenamefont {Fuhrer}, \citenamefont {Schaefer},
  \citenamefont {Faraji}, \citenamefont {Rostami}, \citenamefont {Ghasemi},
  \citenamefont {Sadeghi}, \citenamefont {Grauzinyte},\ and\ \citenamefont
  {Wolverton}}]{zhu2016fingerprint}%
  \BibitemOpen
  \bibfield  {author} {\bibinfo {author} {\bibfnamefont {L.}~\bibnamefont
  {Zhu}}, \bibinfo {author} {\bibfnamefont {M.}~\bibnamefont {Amsler}},
  \bibinfo {author} {\bibfnamefont {T.}~\bibnamefont {Fuhrer}}, \bibinfo
  {author} {\bibfnamefont {B.}~\bibnamefont {Schaefer}}, \bibinfo {author}
  {\bibfnamefont {S.}~\bibnamefont {Faraji}}, \bibinfo {author} {\bibfnamefont
  {S.}~\bibnamefont {Rostami}}, \bibinfo {author} {\bibfnamefont {S.~A.}\
  \bibnamefont {Ghasemi}}, \bibinfo {author} {\bibfnamefont {A.}~\bibnamefont
  {Sadeghi}}, \bibinfo {author} {\bibfnamefont {M.}~\bibnamefont {Grauzinyte}},
  \ and\ \bibinfo {author} {\bibnamefont {Wolverton}},\ }\bibfield  {title}
  {\enquote {\bibinfo {title} {A fingerprint based metric for measuring
  similarities of crystalline structures},}\ }\href@noop {} {\bibfield
  {journal} {\bibinfo  {journal} {J. Chem. Phys.}\ }\textbf {\bibinfo {volume}
  {144}},\ \bibinfo {pages} {034203} (\bibinfo {year} {2016})}\BibitemShut
  {NoStop}%
\bibitem [{\citenamefont {Schaefer}\ and\ \citenamefont
  {Goedecker}(2016)}]{SchaeferComputationally}%
  \BibitemOpen
  \bibfield  {author} {\bibinfo {author} {\bibfnamefont {B.}~\bibnamefont
  {Schaefer}}\ and\ \bibinfo {author} {\bibfnamefont {S.}~\bibnamefont
  {Goedecker}},\ }\bibfield  {title} {\enquote {\bibinfo {title}
  {Computationally efficient characterization of potential energy surfaces
  based on fingerprint distances},}\ }\href {\doibase
  http://dx.doi.org/10.1063/1.4956461} {\bibfield  {journal} {\bibinfo
  {journal} {J. Chem. Phys.}\ }\textbf {\bibinfo {volume} {145}},\ \bibinfo
  {eid} {034101} (\bibinfo {year} {2016})}\BibitemShut {NoStop}%
\bibitem [{\citenamefont {von Lilienfeld}\ \emph {et~al.}(2015)\citenamefont
  {von Lilienfeld}, \citenamefont {Ramakrishnan}, \citenamefont {Rupp},\ and\
  \citenamefont {Knoll}}]{von2015fourier}%
  \BibitemOpen
  \bibfield  {author} {\bibinfo {author} {\bibfnamefont {O.~A.}\ \bibnamefont
  {von Lilienfeld}}, \bibinfo {author} {\bibfnamefont {R.}~\bibnamefont
  {Ramakrishnan}}, \bibinfo {author} {\bibfnamefont {M.}~\bibnamefont {Rupp}},
  \ and\ \bibinfo {author} {\bibfnamefont {A.}~\bibnamefont {Knoll}},\
  }\bibfield  {title} {\enquote {\bibinfo {title} {Fourier series of atomic
  radial distribution functions: A molecular fingerprint for machine learning
  models of quantum chemical properties},}\ }\href@noop {} {\bibfield
  {journal} {\bibinfo  {journal} {Int. J. Quantum Chem.}\ }\textbf {\bibinfo
  {volume} {115}},\ \bibinfo {pages} {1084--1093} (\bibinfo {year}
  {2015})}\BibitemShut {NoStop}%
\bibitem [{\citenamefont {Song}\ and\ \citenamefont
  {Xiao}(2014)}]{song2014sliding}%
  \BibitemOpen
  \bibfield  {author} {\bibinfo {author} {\bibfnamefont {S.}~\bibnamefont
  {Song}}\ and\ \bibinfo {author} {\bibfnamefont {J.}~\bibnamefont {Xiao}},\
  }\enquote {\bibinfo {title} {Sliding shapes for 3d object detection in depth
  images},}\ in\ \href {\doibase 10.1007/978-3-319-10599-4_41} {\emph {\bibinfo
  {booktitle} {Computer Vision -- ECCV 2014: 13th European Conference, Zurich,
  Switzerland, September 6-12, 2014, Proceedings, Part VI}}},\ \bibinfo
  {editor} {edited by\ \bibinfo {editor} {\bibfnamefont {D.}~\bibnamefont
  {Fleet}}, \bibinfo {editor} {\bibfnamefont {T.}~\bibnamefont {Pajdla}},
  \bibinfo {editor} {\bibfnamefont {B.}~\bibnamefont {Schiele}}, \ and\
  \bibinfo {editor} {\bibfnamefont {T.}~\bibnamefont {Tuytelaars}}}\ (\bibinfo
  {publisher} {Springer International Publishing},\ \bibinfo {address} {Cham},\
  \bibinfo {year} {2014})\ pp.\ \bibinfo {pages} {634--651}\BibitemShut
  {NoStop}%
\bibitem [{\citenamefont {Shrivastava}\ and\ \citenamefont
  {Gupta}(2013)}]{shrivastava2013building}%
  \BibitemOpen
  \bibfield  {author} {\bibinfo {author} {\bibfnamefont {A.}~\bibnamefont
  {Shrivastava}}\ and\ \bibinfo {author} {\bibfnamefont {A.}~\bibnamefont
  {Gupta}},\ }\bibfield  {title} {\enquote {\bibinfo {title} {Building
  part-based object detectors via 3d geometry},}\ }in\ \href@noop {} {\emph
  {\bibinfo {booktitle} {The IEEE International Conference on Computer Vision
  (ICCV)}}}\ (\bibinfo {year} {2013})\BibitemShut {NoStop}%
\bibitem [{\citenamefont {Lai}\ \emph {et~al.}(2012)\citenamefont {Lai},
  \citenamefont {Bo}, \citenamefont {Ren},\ and\ \citenamefont
  {Fox}}]{lai2012detection}%
  \BibitemOpen
  \bibfield  {author} {\bibinfo {author} {\bibfnamefont {K.}~\bibnamefont
  {Lai}}, \bibinfo {author} {\bibfnamefont {L.}~\bibnamefont {Bo}}, \bibinfo
  {author} {\bibfnamefont {X.}~\bibnamefont {Ren}}, \ and\ \bibinfo {author}
  {\bibfnamefont {D.}~\bibnamefont {Fox}},\ }\bibfield  {title} {\enquote
  {\bibinfo {title} {Detection-based object labeling in 3d scenes},}\ }in\
  \href {\doibase 10.1109/ICRA.2012.6225316} {\emph {\bibinfo {booktitle}
  {Robotics and Automation (ICRA), 2012 IEEE International Conference on}}}\
  (\bibinfo {year} {2012})\ pp.\ \bibinfo {pages} {1330--1337}\BibitemShut
  {NoStop}%
\bibitem [{\citenamefont {Kim}, \citenamefont {Xu},\ and\ \citenamefont
  {Savarese}(2013)}]{kim2013accurate}%
  \BibitemOpen
  \bibfield  {author} {\bibinfo {author} {\bibfnamefont {B.-s.}\ \bibnamefont
  {Kim}}, \bibinfo {author} {\bibfnamefont {S.}~\bibnamefont {Xu}}, \ and\
  \bibinfo {author} {\bibfnamefont {S.}~\bibnamefont {Savarese}},\ }\bibfield
  {title} {\enquote {\bibinfo {title} {Accurate localization of 3d objects from
  rgb-d data using segmentation hypotheses},}\ }in\ \href@noop {} {\emph
  {\bibinfo {booktitle} {The IEEE Conference on Computer Vision and Pattern
  Recognition (CVPR)}}}\ (\bibinfo {year} {2013})\BibitemShut {NoStop}%
\bibitem [{\citenamefont {Goodfellow}\ \emph {et~al.}(2014)\citenamefont
  {Goodfellow}, \citenamefont {Pouget-Abadie}, \citenamefont {Mirza},
  \citenamefont {Xu}, \citenamefont {Warde-Farley}, \citenamefont {Ozair},
  \citenamefont {Courville},\ and\ \citenamefont
  {Bengio}}]{goodfellow2014generative}%
  \BibitemOpen
  \bibfield  {author} {\bibinfo {author} {\bibfnamefont {I.}~\bibnamefont
  {Goodfellow}}, \bibinfo {author} {\bibfnamefont {J.}~\bibnamefont
  {Pouget-Abadie}}, \bibinfo {author} {\bibfnamefont {M.}~\bibnamefont
  {Mirza}}, \bibinfo {author} {\bibfnamefont {B.}~\bibnamefont {Xu}}, \bibinfo
  {author} {\bibfnamefont {D.}~\bibnamefont {Warde-Farley}}, \bibinfo {author}
  {\bibfnamefont {S.}~\bibnamefont {Ozair}}, \bibinfo {author} {\bibfnamefont
  {A.}~\bibnamefont {Courville}}, \ and\ \bibinfo {author} {\bibfnamefont
  {Y.}~\bibnamefont {Bengio}},\ }\bibfield  {title} {\enquote {\bibinfo {title}
  {Generative adversarial nets},}\ }in\ \href
  {http://papers.nips.cc/paper/5423-generative-adversarial-nets.pdf} {\emph
  {\bibinfo {booktitle} {Advances in Neural Information Processing Systems
  27}}},\ \bibinfo {editor} {edited by\ \bibinfo {editor} {\bibfnamefont
  {Z.}~\bibnamefont {Ghahramani}}, \bibinfo {editor} {\bibfnamefont
  {M.}~\bibnamefont {Welling}}, \bibinfo {editor} {\bibfnamefont
  {C.}~\bibnamefont {Cortes}}, \bibinfo {editor} {\bibfnamefont {N.~D.}\
  \bibnamefont {Lawrence}}, \ and\ \bibinfo {editor} {\bibfnamefont {K.~Q.}\
  \bibnamefont {Weinberger}}}\ (\bibinfo  {publisher} {Curran Associates,
  Inc.},\ \bibinfo {year} {2014})\ pp.\ \bibinfo {pages}
  {2672--2680}\BibitemShut {NoStop}%
\bibitem [{\citenamefont {Denton}\ \emph {et~al.}(2015)\citenamefont {Denton},
  \citenamefont {Chintala}, \citenamefont {szlam},\ and\ \citenamefont
  {Fergus}}]{denton2015deep}%
  \BibitemOpen
  \bibfield  {author} {\bibinfo {author} {\bibfnamefont {E.~L.}\ \bibnamefont
  {Denton}}, \bibinfo {author} {\bibfnamefont {S.}~\bibnamefont {Chintala}},
  \bibinfo {author} {\bibfnamefont {a.}~\bibnamefont {szlam}}, \ and\ \bibinfo
  {author} {\bibfnamefont {R.}~\bibnamefont {Fergus}},\ }\bibfield  {title}
  {\enquote {\bibinfo {title} {Deep generative image models using a
  ￼laplacian pyramid of adversarial networks},}\ }in\ \href
  {http://papers.nips.cc/paper/5773-deep-generative-image-models-using-a-laplacian-pyramid-of-adversarial-networks.pdf}
  {\emph {\bibinfo {booktitle} {Advances in Neural Information Processing
  Systems 28}}},\ \bibinfo {editor} {edited by\ \bibinfo {editor}
  {\bibfnamefont {C.}~\bibnamefont {Cortes}}, \bibinfo {editor} {\bibfnamefont
  {N.~D.}\ \bibnamefont {Lawrence}}, \bibinfo {editor} {\bibfnamefont {D.~D.}\
  \bibnamefont {Lee}}, \bibinfo {editor} {\bibfnamefont {M.}~\bibnamefont
  {Sugiyama}}, \ and\ \bibinfo {editor} {\bibfnamefont {R.}~\bibnamefont
  {Garnett}}}\ (\bibinfo  {publisher} {Curran Associates, Inc.},\ \bibinfo
  {year} {2015})\ pp.\ \bibinfo {pages} {1486--1494}\BibitemShut {NoStop}%
\bibitem [{\citenamefont {Radford}, \citenamefont {Metz},\ and\ \citenamefont
  {Chintala}(2015)}]{radford2015unsupervised}%
  \BibitemOpen
  \bibfield  {author} {\bibinfo {author} {\bibfnamefont {A.}~\bibnamefont
  {Radford}}, \bibinfo {author} {\bibfnamefont {L.}~\bibnamefont {Metz}}, \
  and\ \bibinfo {author} {\bibfnamefont {S.}~\bibnamefont {Chintala}},\
  }\bibfield  {title} {\enquote {\bibinfo {title} {Unsupervised representation
  learning with deep convolutional generative adversarial networks},}\ }\href
  {http://arxiv.org/abs/1511.06434} {\bibfield  {journal} {\bibinfo  {journal}
  {CoRR}\ }\textbf {\bibinfo {volume} {abs/1511.06434}} (\bibinfo {year}
  {2015})}\BibitemShut {NoStop}%
\bibitem [{\citenamefont {Im}\ \emph {et~al.}(2016)\citenamefont {Im},
  \citenamefont {Kim}, \citenamefont {Jiang},\ and\ \citenamefont
  {Memisevic}}]{im2016generating}%
  \BibitemOpen
  \bibfield  {author} {\bibinfo {author} {\bibfnamefont {D.~J.}\ \bibnamefont
  {Im}}, \bibinfo {author} {\bibfnamefont {C.~D.}\ \bibnamefont {Kim}},
  \bibinfo {author} {\bibfnamefont {H.}~\bibnamefont {Jiang}}, \ and\ \bibinfo
  {author} {\bibfnamefont {R.}~\bibnamefont {Memisevic}},\ }\bibfield  {title}
  {\enquote {\bibinfo {title} {Generating images with recurrent adversarial
  networks},}\ }\href {http://arxiv.org/abs/1602.05110} {\bibfield  {journal}
  {\bibinfo  {journal} {CoRR}\ }\textbf {\bibinfo {volume} {abs/1602.05110}}
  (\bibinfo {year} {2016})}\BibitemShut {NoStop}%
\bibitem [{\citenamefont {Yoo}\ \emph {et~al.}(2016)\citenamefont {Yoo},
  \citenamefont {Kim}, \citenamefont {Park}, \citenamefont {Paek},\ and\
  \citenamefont {Kweon}}]{yoo2016pixel-level}%
  \BibitemOpen
  \bibfield  {author} {\bibinfo {author} {\bibfnamefont {D.}~\bibnamefont
  {Yoo}}, \bibinfo {author} {\bibfnamefont {N.}~\bibnamefont {Kim}}, \bibinfo
  {author} {\bibfnamefont {S.}~\bibnamefont {Park}}, \bibinfo {author}
  {\bibfnamefont {A.~S.}\ \bibnamefont {Paek}}, \ and\ \bibinfo {author}
  {\bibfnamefont {I.}~\bibnamefont {Kweon}},\ }\bibfield  {title} {\enquote
  {\bibinfo {title} {Pixel-level domain transfer},}\ }\href
  {http://arxiv.org/abs/1603.07442} {\bibfield  {journal} {\bibinfo  {journal}
  {CoRR}\ }\textbf {\bibinfo {volume} {abs/1603.07442}} (\bibinfo {year}
  {2016})}\BibitemShut {NoStop}%
\bibitem [{\citenamefont {Salimans}\ \emph {et~al.}(2016)\citenamefont
  {Salimans}, \citenamefont {Goodfellow}, \citenamefont {Zaremba},
  \citenamefont {Cheung}, \citenamefont {Radford},\ and\ \citenamefont
  {Chen}}]{goodfellow2016improved}%
  \BibitemOpen
  \bibfield  {author} {\bibinfo {author} {\bibfnamefont {T.}~\bibnamefont
  {Salimans}}, \bibinfo {author} {\bibfnamefont {I.~J.}\ \bibnamefont
  {Goodfellow}}, \bibinfo {author} {\bibfnamefont {W.}~\bibnamefont {Zaremba}},
  \bibinfo {author} {\bibfnamefont {V.}~\bibnamefont {Cheung}}, \bibinfo
  {author} {\bibfnamefont {A.}~\bibnamefont {Radford}}, \ and\ \bibinfo
  {author} {\bibfnamefont {X.}~\bibnamefont {Chen}},\ }\bibfield  {title}
  {\enquote {\bibinfo {title} {Improved techniques for training gans},}\ }\href
  {http://arxiv.org/abs/1606.03498} {\bibfield  {journal} {\bibinfo  {journal}
  {CoRR}\ }\textbf {\bibinfo {volume} {abs/1606.03498}} (\bibinfo {year}
  {2016})}\BibitemShut {NoStop}%
\bibitem [{\citenamefont {Kazachenko}, \citenamefont {Bulusu},\ and\
  \citenamefont {Thakkar}(2013)}]{kazachenko2013methanol}%
  \BibitemOpen
  \bibfield  {author} {\bibinfo {author} {\bibfnamefont {S.}~\bibnamefont
  {Kazachenko}}, \bibinfo {author} {\bibfnamefont {S.}~\bibnamefont {Bulusu}},
  \ and\ \bibinfo {author} {\bibfnamefont {A.~J.}\ \bibnamefont {Thakkar}},\
  }\bibfield  {title} {\enquote {\bibinfo {title} {Methanol clusters (ch3oh) n:
  Putative global minimum-energy structures from model potentials and
  dispersion-corrected density functional theory},}\ }\href@noop {} {\bibfield
  {journal} {\bibinfo  {journal} {J. Chem. Phys.}\ }\textbf {\bibinfo {volume}
  {138}},\ \bibinfo {pages} {224303} (\bibinfo {year} {2013})}\BibitemShut
  {NoStop}%
\end{thebibliography}%

\end{document}